\documentclass[longauth]{aa}  

\usepackage{graphicx}
\usepackage{txfonts}

\newcommand{\heimdall}{\texttt{Heimdall} }

\newcommand{\pcm}{\,pc\,cm$^{-3}$}

\usepackage[colorlinks=true,citecolor=blue]{hyperref}
\begin{document}

   \title{Simultaneous and panchromatic observations of the Fast Radio Burst FRB\,20180916B}
   \titlerunning{Panchromatic Observations of FRB\,20180916B}
   \subtitle{}

   \author{M. Trudu \inst{1,2} 
          \and
          M. Pilia \inst{2}
          \and 
          L. Nicastro \inst{3}
          \and 
          C. Guidorzi \inst{4,3,5}
          \and 
          M. Orlandini \inst{3}
          \and 
          L. Zampieri \inst{6}
          \and 
          V. R. Marthi \inst{7}
          \and 
          F. Ambrosino \inst{8,9}
          \and           
          A. Possenti \inst{2,1}
          \and 
          M. Burgay \inst{2}
          \and 
          C. Casentini \inst{8,10}
          \and 
          I. Mereminskiy \inst{11}
          \and 
          V. Savchenko \inst{12}
          \and 
          E. Palazzi \inst{3}
          \and 
          F. Panessa \inst{8}
          \and 
          A. Ridolfi \inst{2}
          \and 
          F. Verrecchia \inst{13,14}
          \and 
          M. Anedda \inst{1}
          \and 
          G. Bernardi \inst{15,16,17}
          \and 
          M. Bachetti \inst{2}
          \and 
          R. Burenin \inst{11}
          \and 
          A. Burtovoi \inst{18}
          \and 
          P. Casella \inst{14}
          \and 
          M. Fiori \inst{19}
          \and 
          F. Frontera \inst{4}
          \and 
          V Gajjar \inst{20}
          \and 
          A. Gardini \inst{21,22}
          \and 
          M. Ge \inst{23}
          \and 
          A. Guijarro-Rom\'an \inst{22}
          \and 
          A. Ghedina \inst{24}
          \and 
          I. Hermelo \inst{22}
          \and 
          S. Jia \inst{23}
          \and 
          C. Li \inst{23}
          \and 
          J. Liao \inst{23}
          \and 
          X. Li \inst{23}
          \and 
          F. Lu \inst{23}
          \and 
          A. Lutovinov \inst{11}
          \and 
          G. Naletto \inst{19,6}
          \and 
          P. Ochner \inst{19,6}
          \and 
          A. Papitto \inst{4}
          \and 
          M. Perri \inst{13,14}
          \and 
          C. Pittori \inst{13,14}
          \and 
          B. Safonov \inst{25}
          \and 
          A. Semena \inst{11}
          \and 
          I. Strakhov \inst{25}
          \and 
          M. Tavani \inst{8,26}
          \and 
          A. Ursi \inst{8}
          \and 
          S. L. Xiong \inst{23}
          \and 
          S. N. Zhang \inst{23,27}
          \and 
          S. Zheltoukhov \inst{25}
          }

   \institute{Università degli Studi di Cagliari, Dipartimento di Fisica, SP Monserrato-Sestu km 0.7, I-09042 Monserrato  (CA), Italy
    \and
    INAF-Osservatorio Astronomico di Cagliari, via della Scienza 5, I-09047, Selargius (CA), Italy
    \and 
    INAF - Osservatorio di Astrofisica e Scienza dello Spazio di Bologna, Via Piero Gobetti 93/3, I-40129 Bologna, Italy
    \and 
    Department of Physics and Earth Science, University of Ferrara, via Saragat 1, I-44122 Ferrara, Italy
    \and 
    INFN – Sezione di Ferrara, via Saragat 1, I-44122 Ferrara, Italy
    \and 
    INAF - Osservatorio Astronomico di Padova - Vicolo dell’Osservatorio 5 - 35122 Padova, Italy
    \and 
    National Centre for Radio Astrophysics, Tata Institute of Fundamental Research, Post Bag 3, Ganeshkhind, Pune - 411 007, India
    \and 
    INAF/IAPS, via del Fosso del Cavaliere 100, I-00133 Roma (RM), Italy
    \and 
    Sapienza Universit\'a  di Roma, Piazzale Aldo Moro 5, I-00185 Roma (RM), Italy
    \and 
    INFN Sezione di Roma 2, via della Ricerca Scientifica 1, I-00133 Roma (RM), Italy
    \and 
    Space Research Institute, Russian Academy of Sciences, Profsoyuznaya 84/32, 117997 Moscow, Russia
    \and 
    ISDC, Department of Astronomy, University of Geneva, Chemin dcogia, 16 CH-1290 Versoix, Switzerland
    \and 
    SSDC/ASI, via del Politecnico snc, I-00133 Roma (RM), Italy
    \and 
    INAF - Osservatorio Astrofisico di Roma, via Frascati 33, I-00078 Monte Porzio Catone (RM), Italy
    \and 
    INAF-Istituto di Radio Astronomia, via Gobetti 101, 40129 Bologna, Italy
    \and 
    Department of Physics and Electronics, Rhodes University, PO Box 94, Grahamstown, 6140, South Africa
    \and 
    South African Radio Astronomy Observatory, Black River Park, 2 Fir Street, Observatory, Cape Town, 7925, South Africa
    \and 
    INAF - Osservatorio Astrofisico di Arcetri, Largo Enrico Fermi 5, 50125, Florence, Italy
    \and 
    Department of Physics and Astronomy, University of Padova, Via F. Marzolo 8, 35131, Padova, Italy
    \and 
    Department of Astronomy, University of California Berkeley, Berkeley CA 94720
    \and 
    Instituto de Astrofísica de Andalucía, Glorieta de la Astronomía s/n, 18008 Granada, Spain
    \and 
    CAHA - Centro Astron\'omico Hispano en Andaluc\'\i a, Observatorio de Calar Alto, Sierra de los Filabres, 04550 G\'ergal, Almer\'ia, Spain
    \and 
    Key Laboratory of Particle Astrophysics, Institute of High Energy Physics, Chinese Academy of Sciences, 19B Yuquan Road, Beijing 100049, PR China
    \and 
    Fundaci\'on Galileo Galilei - INAF - Rambla J.A.Fern\'andez P. 7, 38712, S.C.Tenerife, Spain
    \and 
    Sternberg Astronomical Institute, Moscow M.V. Lomonosov State University, Universitetskij pr., 13, 119992, Moscow, Russia
    \and 
    Universit\'a  degli Studi di Roma ”Tor Vergata”, via della Ricerca Scientifica 1, I-00133 Roma (RM), Italy
    \and 
    University of Chinese Academy of Sciences, Chinese Academy of Sciences, Beijing 100049, PR China
             }
   \date{Received September 15, 1996; accepted March 16, 1997}

 
  \abstract
   {}
   {Fast Radio Bursts are bright radio transients whose origin has not yet been explained. The search for a multi-wavelength counterpart of those events can put a tight constrain on the emission mechanism and the progenitor source. }
   {We conducted a multi-wavelength observational campaign on FRB\,20180916B between October 2020 and August 2021 during eight activity cycles of the source. Observations were led in the radio band by the SRT both at 336 and 1547 MHz and the uGMRT at $400$\,MHz. Simultaneous observations have been conducted by the optical telescopes Asiago (Galileo and Copernico), CMO SAI MSU, CAHA 2.2m, RTT-150 and TNG, and X/$\gamma$-ray detectors on board the AGILE, \emph{Insight--\/}HXMT, INTEGRAL and Swift satellites. } 
   {We present the detection of 14 new radio bursts detected with the SRT at 336\,MHz and seven new bursts with the uGMRT from this source. We provide the deepest prompt upper limits in the optical band for FRB\,20180916B to date. In fact, the TNG/SiFAP2 observation simultaneous to a burst detection by uGMRT gives an upper limit $E_{{\rm optical}} / E_{{\rm radio}}  < 1.3 \times 10^{2}$. Another burst detected by the SRT at $336$\,MHz was also co-observed by \emph{Insight--\/}HXMT. The non-detection in the X-rays yields  an upper limit ($1-30$\,keV band) of $E_{{\rm X-ray}} / E_{\text{radio}}$ in the range of $(0.9-1.3) \times 10^7$, depending on which model is considered for the X-ray emission.}
   {}

   \keywords{methods: observational -- Instrumentation: photometers -- X-ray: bursts -- stars: magnetars -- stars: neutron -- stars: flare}

   \maketitle
%

\section{Introduction}
\label{sec:introduction}
Fast Radio Bursts \citep[FRBs, for a review see e.g.][]{petroff_2022_frbreview} are intense ($10^{36-40}$\,erg), radio flashes of millisecond-duration coming from extragalactic distances. The origin behind these astrophysical transients is still under debate. Most of them appear to be sporadic single events and only a small percentage shows a repeating behaviour \citep[see e.g.][]{spitler16, chime_2019_8newfrb}.

Discovered by the CHIME telescope as the third known repeater \citep{chime8_19}, FRB\,20180916B (R3) is the first FRB for which periodic activity was detected \citep{chime_2020_periodr3}. It has a period of $16.33\pm 0.12$\,days \citep{pleunis21a, chime_2020_periodr3} with an active window of $\pm 2.6$\,days as observed at CHIME's frequencies: 400--800 MHz. Soon afterwards, periodic activity was also claimed for the first discovered repeater, FRB\,20121102A \citep[R1, ][]{spitler16,spitler14}, with an activity period of about 160\,days and a duty cycle of $\sim 54$\,\% \citep{cruces21,rajwade20}.

R3 detections at different radio frequencies, ranging from $\sim 150$ MHz \citep{pastor-marazuela_2021_r3, pleunis21a} to 6 GHz \citep{bethapudi22}, indicate a chromatic pattern in the burst occurrence. The bursts at higher and lower frequencies than CHIME only partially overlap with the CHIME active phase window. Specifically, bursts detected at 1.4 GHz, such as those detected by Apertif \citep{pastor-marazuela_2021_r3}, have an occurrence distribution shifted $\sim 0.7$ days earlier (i.e., earlier activity phases) compared to the CHIME detected ones. In contrast, bursts detected at the lower LOFAR frequencies show a phase shift of $\sim 3$ days (i.e., later activity phases) \citep{pleunis21a}. The differences in the active window include not only the different onset times but also the duration: the higher frequency active phases seem to last shorter than the lower frequency ones \citep{bethapudi22, pastor-marazuela_2021_r3}. Meanwhile, the average width of the bursts becomes larger at lower frequencies \citep{pastor-marazuela_2021_r3}. Although scattering contributes to this effect, it is generally limited to only the lowest frequencies for this FRB \citep{marcote20, pilia20, pastor-marazuela_2021_r3}.

Although the FRBs origin is yet an open question, most of the current models consider highly magnetised neutron stars, i.e. magnetars, as one of the most favoured FRB progenitor \citep[][]{margalit_2020_frbmagnetar, beloborodov_2017_r1magnetar,lyubarsky14}. The magnetar models predict the FRB event to happen either via magnetic reconnection in the neutron star magnetosphere or via shock(s) due to the interaction of a powerful magnetar outflow with the surrounding medium \citep[see e.g.][for a comprehensive review]{zhang_2022_frbreview}. The FRB-magnetar link  was strengthened with the detection of a radio burst similar to an FRB from the Galactic magnetar SGR\,J1935$+$2154 \citep{chime20sgr,bochenek20} accompanied by simultaneous X-ray emission \citep{mereghetti20,Tavani2021,hxmt_sgr,ridnaia20}.  Whether magnetars or other sources (e.g. black holes, ultra-luminous X-ray sources) are considered as progenitors, multi-wavelength (MWL) emission is predicted by most models, in the form of prompt or afterglow \citep[see e.g.][]{beloborodov_2020,lyutikov_2020_frbobstars,metzger2019,ghisellini18,kumar17,beloborodov_2017_r1magnetar,lyubarsky14}. The detection of MWL emission from an FRB source would be an important piece of information to discriminate amongst the current models. 

A deep search for MWL emission has been performed in the early days for FRB\,20140514A \citep{petroff_2015_frbmwl} and more than once for R1 \citep[][]{hardy_2017_r1,scholz_2017_r1mwl, scholz_2016_r1mwl}.

The search for simultaneous emission at other wavelengths is better targeted in the case of a repeater, where one can expect to observe bursts during a MWL campaign. R1, however, is relatively far \citep[$\sim1$\,Gpc,][]{chatterjee17,marcote17} and this made the chances of detection with current optical/high-energy instruments slim. On the other hand R3, with its proximity ($\sim$ 149 Mpc, \citealt{marcote20}) and, furthermore, its periodic repetition, is a more suitable candidate for MWL coordinated observational campaigns, and indeed several were performed \citep{pearlman20, pilia20, scholz20}, however reporting no MWL detection.

In this work we present a MWL observational campaign on R3 that has been conducted in the radio band with the Sardinia Radio Telescope (SRT) and the upgraded Giant Metrewave Radio Telescope (uGMRT) over the course of eight activity cycles of the source. Optical to $\gamma$-ray telescopes were used to shadow the radio observations and seek for higher energy emission. We report the detection of 21 radio bursts from the source and a null detection at other wavelengths, providing for the optical band and the high-energies stringent upper limits (ULs). The paper is organised as follows: Sec.\,\ref{sec:obs} presents an overview of the instruments involved and the observing times and tools; Sec.\,\ref{sec:results} presents the detections in the radio band and the ULs at other wavelengths; in Sec.\,\ref{sec:discussion} we discuss our results; in Sec.\,\ref{sec:conclusion} we provide our conclusions.

\section{Observations and Data Reduction} \label{sec:obs}

\begin{figure*}
    \centering
    \includegraphics[width=\linewidth]{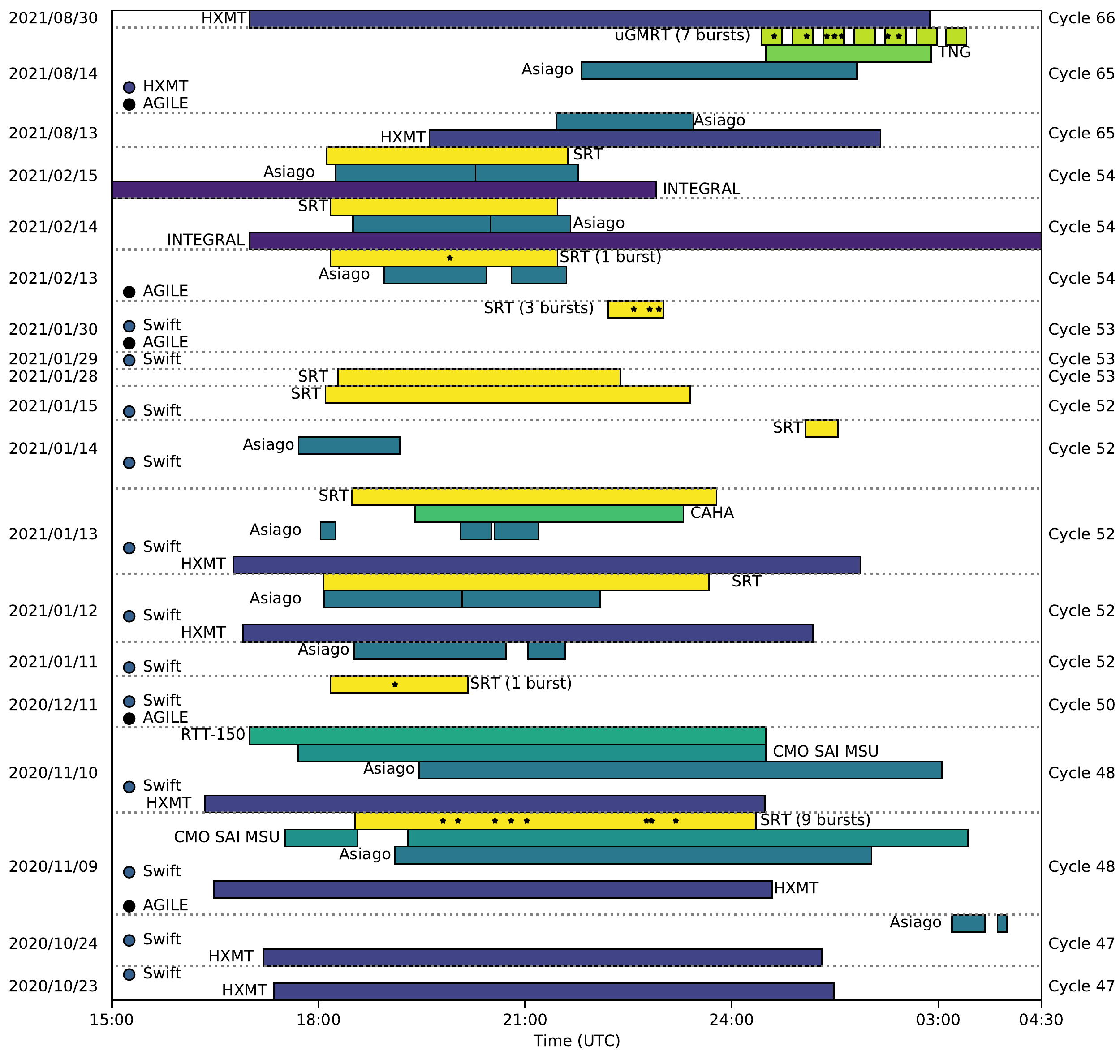}
    \caption{Observational R3 MWL campaign. Coloured bars represent observations performed for each facility as a function of time of the day (UTC). For each day (reported on the left) the corresponding activity cycle number of R3 is reported on the right. The time after 24:00 should be considered as part of the day after. Each detected radio burst is reported as a black star. \emph{AGILE} and Swift (and \emph{Insight--\/}HXMT on 2021/08/14) were observing throughout the whole day, except during epochs of visibility limitations (Earth occultations and South Atlantic Anomaly passages). The days in which a single instrument is present are scheduled days in which observations with other instruments have been cancelled.}
    \label{fig:campaign} 
\end{figure*}

We performed a MWL observing campaign on R3, focusing on different activity cycles, from 2020/10/23 to 2021/08/30. In this section, for each wavelength, we report the specifications of the instruments used and how the data have been processed. The whole campaign, with all the telescopes involved and the observing times, is summarised in Fig.\ref{fig:campaign}. 

\subsection{Radio Instruments} \label{subsec:obsradio}

\begin{table*}
\caption{Details of the radio instruments involved in the R3 MWL campaign. For each instrument we report the central frequency $\nu_c$, observational bandwidth BW, sampling time $d t$, spectral resolution $d \nu$, system equivalent flux density (SEFD) and the fluence density\tablefootmark{a} threshold (S/N=6) $F_{\nu}^{\rm min}$. $N_A$ is the number of antennas deployed with the uGMRT array.}  
\label{tab:radiotelescopes}
\centering                          
\begin{tabular}{l l c c c c c c}        
\hline\hline                 
Telescope & Receiver & $\nu_c$ & BW   & $d t$ & $d \nu$ & SEFD & $F_{\nu}^{\rm min}$\\ 
          &          & $\mathrm{[MHz]}$ & $\mathrm{[MHz]}$ & $\mathrm{[\mu s]}$ & $\mathrm{[kHz]}$ & $\mathrm{[Jy]}$ & $\mathrm{[Jy\,ms]}$\\
\hline                        
SRT & L/P L-Band & 1547 & 512 & 125 & 1000 & 37  & 0.4 \\ 
    & L/P P-Band & 336 & 80   & 128 & 250  & 215 & 4.6 \\ 
uGMRT & Band-3 & 400 & 200 & 81.92 & 97.65 & (502/$N_A$)& 0.6 \\ 
\hline                                   
\end{tabular}
\tablefoot{
\tablefoottext{a}{Throughout this work, we refer to the source irradiated energy per surface unit as ``fluence'', and to the fluence per frequency unit as ``fluence density''.}

}
\end{table*} 

\subsubsection{SRT} \label{subsubsec:srt}

The SRT monitored R3 between 2020/11/09 and 2021/02/15, for a total observing time of $\sim 39$\,hours. The observations were made with the L/P dual band receiver \citep[SRT-L and SRT-P, ][]{valente+10}, with observational central frequency of 1547 and 336 MHz respectively (see Tab.\,\ref{tab:radiotelescopes}). The L-band data were recorded with the ATNF Digital Filterbank Mark III backend (DFB) whereas the P-band data were acquired via the ROACH1 backend \citep{bassa+16}. 

The L-band data were recorded as 2-bits per sample {\sf psrfits} \citep{psrchive} search-mode files and subsequently converted into 8-bits {\sf SIGPROC} \citep{sigproc} filterbank files; the P-band, after being recorded as {\sf dada} baseband files, are coherently dedispersed at dispersion measure ${\rm (DM)} = 348.82$\,\pcm\ \citep{chime_2019_8newfrb}, to remove  the intra-channel smearing, and converted into 8-bits filterbank files using the {\sf DSPSR} \citep{dspsr} routine {\sf digifil}. The L-band receiver of 512\,MHz bandwidth is divided into frequency channels of 1\,MHz each and time sampled at 125\,$\mu$s. The L-band data are originally recorded with a larger bandwidth of 1024\,MHz to avoid aliasing effects. Radio frequency interference (RFI) typically affects less than 20\,\% of the nominal 512\,MHz band. The $80$\,MHz bandwidth of the P-band is split into 320 channels of $0.25$\,MHz each and sampled at 128\,$\mu$s. The P-band bandwidth channels are generally corrupted by RFI for about 30\,\%.



\subsubsection{uGMRT} \label{subsubsec:ugmrt}
The uGMRT \citep{Gupta-uGMRT} is an enhancement of the GMRT \citep{Swarup-GMRT} with improved receivers providing nearly seamless coverage from $50$ to $1500$\,MHz and wide instantaneous bandwidths. The observations were carried out with the Band-3 receiver in the frequency range $300-500$\,MHz. The GMRT Wideband Backend \citep[GWB,][]{Reddy-GWB} was configured to split the $200$\,MHz bandwidth into 2048 spectral channels, giving a channel width of $97.65$\,kHz. At the time of the observation, the signals from 27 available antennas were combined in phase to give a phased array beam which was coherently dedispersed at ${\rm DM} = 348.82$\,\pcm\ . The real-time coherent dedispersion corrects only for the intra-channel dispersion, while the inter-channel dispersion is expected to be corrected offline. The array was rephased every 20 minutes to undo the phase drift due to ionospheric effects on the far away arm  antennas. The calibrator 3C48 was observed for a few minutes initially to provide the flux scale for fluence calibration in case of bursts detection. Custom total intensity raw data produced by the backend are then converted into 16-bits {\sf SIGPROC} filterbanks\footnote{https://github.com/alex88ridolfi/ugmrt2fil}.
The uGMRT observed R3 on 2021/08/15 for a total observing time of $\sim 2.1$\,hours. 

\subsubsection{Single Pulse Search} \label{subsubsec:spsearch}

The search for radio bursts from R3 was performed using a search pipeline based on \heimdall \citep{bbb+12}. As a first step, an RFI excision is made using the spectral kurtosis algorithm \citep{nita2020} provided by the FRB software package {\sf YOUR} \citep{your}. The DM search was restricted to the range 300--400 \pcm. For the SRT-P data we chose a signal-to-noise ratio (S/N) loss tolerance in each DM trial of 1\%. For the SRT-L and uGMRT data we performed a sub-banded search \citep[similarly to][]{kumarsub}. In the case of SRT-L,
the sub-banded search considered the full 1024\,MHz dataset. In this way the search for bursts was performed within the nominal $1300-1800$\,MHz, but we also took advantage of the fact that the receiver passband filter does not have a sharp drop at its edges and a non-negligible fraction of the signal of a putative band-limited burst could appear outside the nominal borders of the band.
We perfomed the sub-banded search by applying the following frequency windows: we took $[512, 256]$\,MHz windows (for each sub-band we took the overlapping adjacent sub-bands by shifting the bands by half widths) and we used a S/N tolerance of $[1, 0.1]$\,\% with respect to the previously mentioned sub-band widths. For the uGMRT data we considered windows of $[200,100]$\,MHz and for each window a tolerance of $[1,0.5]$\,\%. In all the searches we used a maximum boxcar width of $500$\,ms.


In order to filter the high number of candidates found by \heimdall, we considered some thresholds to sift them using the code \texttt{frb\textunderscore detector.py} \citep{bbb+12}. We selected only candidates with S/N $\geq 6$, and a minimum number of members (distinct boxcar/DM trial) clustered into a single candidate by \heimdall of 10. Lastly, all the candidates were visually inspected. 

\subsection{Optical Instruments} \label{subsec:obsoptical}

\subsubsection{Asiago}\label{asiago}

Extensive optical coverage during periods of radio activity was attained with the fast photon counters Aqueye+ and IFI+Iqueye mounted at the Copernicus and Galileo telescopes, respectively, in Asiago \citep{zampieri19,zampieri15, naletto_2009_iqueye, barbieri_2009_aqueye}. The two instruments performed several observations of the area of the sky centered at the position of R3 \citep[R.A. = $01^{\rm h}58^{\rm m}00^{\rm s}.75017 \pm 2.3$ mas, Dec. = $65^\circ 43^\prime 00^{\prime\prime}.3152 \pm 2.3$ mas,][]{marcote20} at the dates reported in Tab. \ref{tab:asiago} (some of them are simultaneous). Observations were performed in white light. At the beginning and at the end of each acquisition, a reference star close to the position of the target (2MASS 01580548+6542269, offset 44") was also observed to monitor the quality and transparency of the sky, i.e. to check that the count rate was constant and consistent with the expected value (according to the calibration relations in \citealt{zampieri16}). The same star (for Aqueye+) or another field star (2MASS 01580026+6541437, for IFI+Iqueye) was selected and carefully centred on the instrument camera for guiding purposes, in such a way that the FRB position matches the aperture of the on-source detectors or the fiber position. To this aim, an image of the field was previously acquired and astrometrically calibrated. The error of the target position registered on the image is $\sim 0.6^{\prime\prime}$, significantly smaller than the aperture/fiber diameter.

The photon event lists were reduced using a dedicated software (QUEST v. 1.1.5, \citealt{zampieri15}). Light curves with different bins widths were computed from the reduced event lists and searched for any significant rate increase. Periods of bad sky quality (count rate of the on-source detectors $> 3700$\,counts/s for Aqueye+ and $> 2600$\,counts/s for IFI+Iqueye) were discarded. The total useful on-source observing time was  $\sim 129.5$\,ks for Aqueye+ and $\sim 62.1$\,ks for IFI+Iqueye (data discarded $\sim 10-20$\%). The count rate averaged over all acquisitions was $3038$\,counts/s for Aqueye+ and $1282$\,counts/s for IFI+Iqueye.

\subsubsection{CAHA 2.2m}
The Calar Alto 2.2m telescope equipped with the AstraLux camera observed R3 for $3.3$\,hours starting on 2021/01/13 19:24:25 UT. A GG 385 longpass filter was used (White $\sim 667 \pm 237$ nm).
AstraLux \citep{Hormuth08} uses an electron-multiplying, thinned, and back-illuminated $512 \times 512$ pixel CCD manufactured by Andor.
It provides a field of view of $24'' \times 24''$ at a pixel scale of 47 mas/pixel. Using subarrays, binning, and short vertical shift times allows frame rates of more than 1 kHz. Operated as electron-multiplying CCD, as in the case of our observation, AstraLux can reach multiplication gains of up to 2500 with a linear response up to 70 photons per pixel.

\begin{figure}
    \centering
    \includegraphics[width=\linewidth]{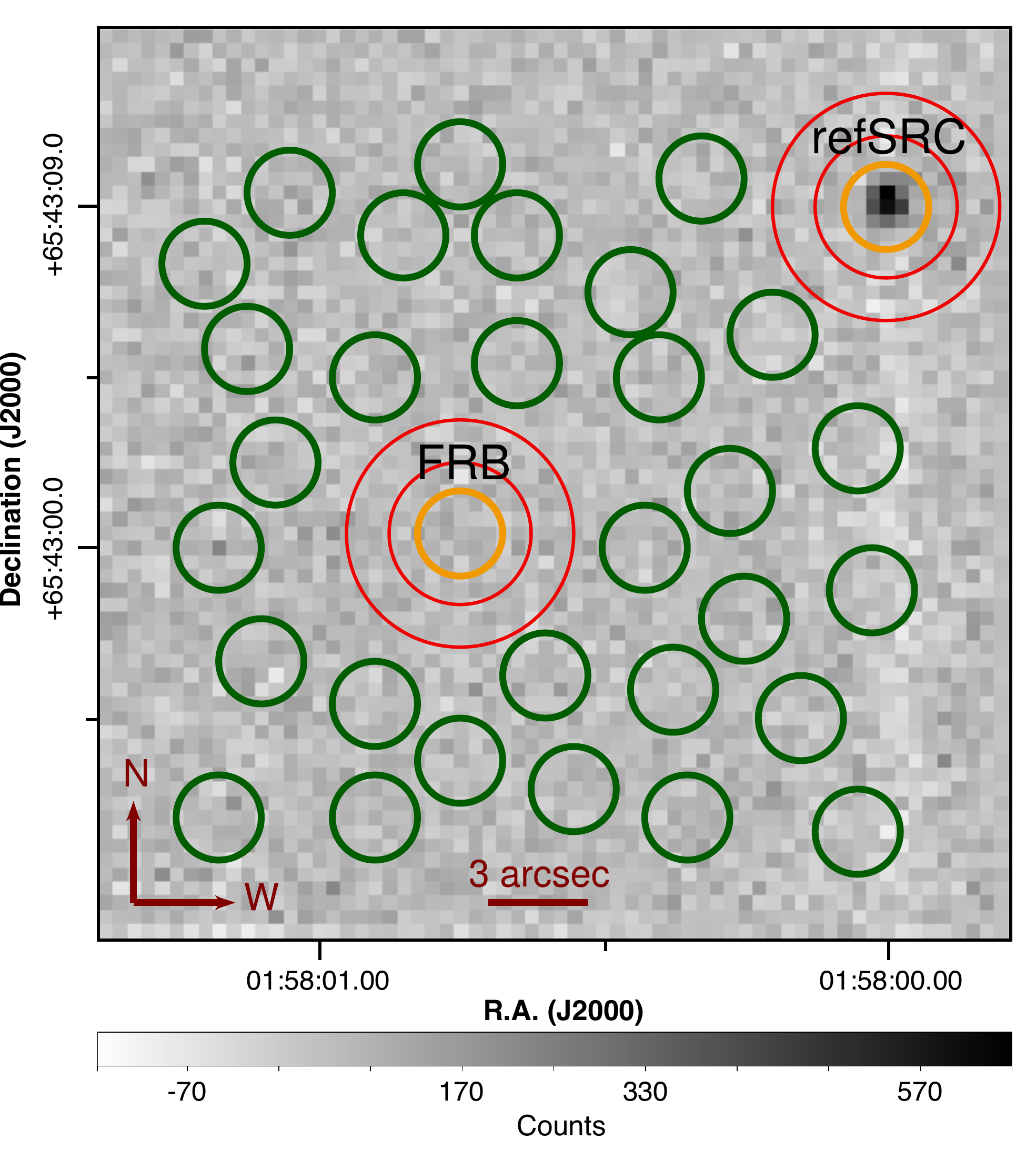}
    \caption{Sum of twenty 5-ms frames collected by AstraLux. The R3 position and the reference field source are marked by orange circles. The rings where the background was computed are marked in red. In green the thirty random regions whose aperture photometry and count statistics \textcolor{blue}{were} investigated.}
    \label{fig:AstraLux_regions}
\end{figure}

Our observation was performed applying a $8 \times 8$ binning ($0.376^{\prime\prime}$/pixel) and acquisition rate of $200$\,Hz, which gives $64 \times 64$ bin frames, $5$\,ms integration time each. 
Split in 4 observing blocks due to the constraints of the acquisition software, we accumulated a total of 2,360,000 frames.
In order to include in the field of view the G $\simeq 15.6$ reference star located North-West of the FRB position, an offset of $\delta {\rm RA} \sim 8.8''$ and $\delta {\rm Dec} \sim -7.3''$, with respect to the star, was applied to the pointing (see Fig.\,\ref{fig:AstraLux_regions}). The post-processing centre coordinates resulted to be R.A. = $01^{\rm h}58^{\rm m}00^{\rm s}.60$, Dec. = $65^\circ 43^\prime 01^{\prime\prime}.4$.

Custom software written in the IDL language was used to manage and analyse the custom AstraLux files. The procedure to identify a possible source onset consisted in aperture photometry at the FRB position, the mentioned reference field source and thirty additional ``void'' regions, as shown in Fig.~\ref{fig:AstraLux_regions}. A three pixel extraction radius ($\sim 1.2''$) was adopted. Pointing drift was accounted for by monitoring the position of the reference source using 1-s accumulated images. The background was monitored in regions around the given positions (area of an external corona delimited by red circles in Fig.\,\ref{fig:AstraLux_regions}) whereas the bias level was computed from the average of the lower 12 rows of each frame to account for a variable, columns dependent variability. Frames time tagging precision is of the order of 0.1 s. 


\subsubsection{2.5 m telescope CMO SAI MSU and 1.5 m RTT-150}
We observed the position of R3 with the 2.5~m telescope of the Caucasian Mountain Observatory (CMO) of the Sternberg Astronomical Institute (SAI) Lomonosov Moscow State University (MSU). We used the Andor iXon 897 camera mounted on Cassegrain focus. This allowed us to obtain 4 ms frames (with 0.28 ms readout time). No optical filter was used in order to maximise the sensitivity. Due to software limitations, the time tag was rounded to 1 s, thus introducing a systematic error in the absolute timing. We selected a $80^{\prime\prime}\times 20^{\prime\prime}$ field with two bright comparison stars with mean Gaia G magnitudes of 13.9 and 15.6: Gaia DR2 518385480138840704 and Gaia DR2 518386236053079936, respectively. The latter corresponds to the reference source observed by AstraLux. The angular scale after binning was $0.32^{\prime\prime}$/pixel.

We obtained about 14 hours of useful exposure over the nights of 2020/11/09 and 2020/11/10, with single series duration of 1200 s. For each individual series an averaged image was produced and astrometric corrections, with respect to known Gaia stars, were performed. Then for each single frame of the series we once again corrected the position with respect to the brightest star in the field (which produced about 750 electrons above the noise level per frame). This allowed us to obtain a typical accuracy of the astrometric solution of $0.2^{\prime\prime}$.

Over the second night of 2020/11/10 we also performed simultaneous observations with the Andor iXon 888 camera mounted on the 1.5~m Russian-Turkish Telescope (RTT-150), accumulating about 6 hours of data. No optical filter was used. We chose to select a single frame exposure of 8.24~ms with 0.1~ms readout time. The same two Gaia stars were used for astrometry. The expected noise signal distribution was acquired from the empty regions with no apparent stars in the field.

\subsubsection{3.6\,m TNG}\label{TNG}
We carried out a single observation of R3 with the fast optical photometer SiFAP2 \citep{Ghedina2018, ambrosino16} mounted at 3.6\,m INAF's Telescopio Nazionale Galileo (TNG) located at Roque de Los Muchachos (La Palma, Canary Islands) on 2021/08/15. The observation was performed without any interruption starting at 00:35:00.000\,UT and lasting 8.7\,ks. No filter was used during the run to maximise the photon count rate. The final position of R3 (see Sec.\,\ref{asiago}) was reached by offsetting the telescope from GSC2.2 N31321329057 star (R.A. = $01^{\rm h}58^{\rm m}05^{\rm s}.50$, Dec. = $65^\circ 42^\prime 26^{\prime\prime}.87 $) with a pointing error $< 0''.5$, significantly better than the Field of View of SiFAP2 ($7'' \times 7''$). The sky background located $\sim 3.5'$ from the target was also monitored simultaneously with R3.

\subsection{High-Energy Instruments} \label{subsec:obshighenergy}

\subsubsection{\emph{AGILE}} \label{subsubsec:AGILE}

The \emph{AGILE} satellite observed the sky region in which R3 is located with its three detectors. The $\gamma$-ray imaging detector (GRID), sensitive in the range 30 MeV -- 30 GeV with a $2.5$\,sr field of view (FoV), a coded mask X-ray imager, Super-AGILE \citep[Super-A; operating in the energy range 18\,--\,60 keV;][]{Feroci2007} with $\sim$\,1\,sr FoV, and the Mini-Calorimeter (MCAL), sensitive in the 0.4--100 MeV band with $4\pi$ non-imaging acceptance \citep{Tavani2009, Tavani2019}.
\emph{AGILE} is currently operating in spinning mode, with the instrument axis rotating every $\sim 7$ minutes around the satellite-Sun direction. For each satellite revolution, a large fraction of the sky ($\sim 40-60$\%) is exposed, depending on the Earth occultation pattern and trigger disabling over the South Atlantic anomaly (SAA, about $10$\% of the 95-min orbit). In particular, on timescales of hours $\sim$80\% of the entire sky can be exposed by the GRID $\gamma$-ray imager and by the MCAL.

\emph{AGILE} has been involved in the MWL campaigns for FRBs since 2019, particularly for R3 \citep{tavani20, Casentini2020}. The advantage to use an instrument like \emph{AGILE} is closely related to its currently operating mode, the spinning. We are able to follow an FRB source minutes before and after each specific burst time with a small temporal gap. 
No evidence of $\gamma$-ray emission from these sources has been found so far \citep{Verrecchia2021, Tavani2021}.
AGILE-GRID exposure near burst times has been checked on short timescales ($\pm\,100$\,s around the bursts), and resulted in a partial coverage of the 21 bursts reported (only 12), due to SAA passages or Earth occultation. The long-timescale GRID analysis has been performed using the standard AGILE multi-source maximum likelihood \citep[AML;][]{Bulgarelli2012}, usually applied to exposures longer than a few hours. We applied standard cuts to events, excluding SAA passages time intervals and events at off-axis angles greater than 60\,$\deg$ or at angles from Earth direction smaller than 80\,$\deg$.

\subsubsection{\emph{Insight--\/}HXMT} \label{subsubsec:HXMT}

\emph{Insight--\/}HXMT is China's first X--ray astronomy mission \citep{2020SCPMAoverview}. It is equipped with three main X--ray instruments operating in the broad 1--250~keV band: the Low Energy (LE) X--ray telescope covers the 1--15 keV band \citep{2020SCPMALE}; the Medium Energy (ME) instrument covers the 5--30~keV band \citep{2020SCPMAME}; and the 20--250~keV band is covered by the High Energy (HE) instrument \citep{2020SCPMAHE}. The time resolution of the LE, ME, and HE are 980\,$\mu$s, 255\,$\mu$s, and less than 10\,$\mu$s, respectively.  The systematic errors of the timing system are 15.8, 8.6, and 12.1 $\mu$s, for LE, ME, and HE, respectively \citep{Tuo2022}.

While the general procedure we followed to extract the light curves is the same as detailed in \cite{Guidorzi2020}, in order to increase the exposure coverage of the observations we lifted some of the standard constraints employed to determine the good time intervals (Li C.-K. and Ji Long, personal communication). In particular, the \emph{only} constraints passed to the \texttt{legtigen}, \texttt{megtigen} and \texttt{hegtigen} tasks were: Earth elevation angle $\text{ELV}>1$~deg for LE and ME, $\text{ELV}>0^\circ$ for HE; the pointing offset angle $\text{ANG\_DIST} \leq 0.5^\circ$ for HE.  For all the three instruments the SAA\_FLAG (exclude the data in the South Atlantic Anomaly) was set to false.

The effect of this non-standard procedure is to not filter the data for high background regions. Because our pipeline estimates the background independently \citep{Guidorzi2020}, this does not affect our results. The advantage, on the other hand, is to cover a much larger time exposure, therefore increasing significantly the chance to find bursts.

The \emph{Insight--\/}HXMT data analysis was performed with the software package HXMTDAS v2.04 and CALDB v2.05. We used the \texttt{hxbary} to apply the barycentric correction, thus converting the photons' arrival terrestrial times (TTs) to the corresponding barycentric dynamic times (TDBs). LE, ME and HE instruments total on source exposure was of 127.7, 118.8 and 80.3 ks respectively. Table~\ref{tab:Xobslog} reports the observation log.

\subsubsection{INTEGRAL} \label{subsubsec:INTEGRAL}

In total, between October 2020 and August 2021, INTEGRAL observed R3 for 658.0~ks (8.0~days) in 276 pointings. All of these data are currently public. INTEGRAL observations consist of approximately 1-hour-long pointings (binned at 1\,s resolution),  with a few hundred seconds slews in between. For the purposes of the observation log, we group them into longer observations, as long as the pointings are separated by no more than 300~seconds, as reported in Table~\ref{tab:integralobs}. The INTEGRAL data analysis has been conducted by using the standard INTEGRAL Offline Scientific Analysis software (version 11.0).  

\subsubsection{Swift} \label{subsubsec:Swift}

The Neil Gehrels Swift Gamma-ray Burst Observatory (Swift) observed R3 with the X-ray Telescope \citep[XRT, ][]{burrows_2005_swiftxrt}, 
one of the three instruments equipped in the facility. The Swift/XRT data were obtained within the planned coordinate multi-instrument campaign started in 2020 whose first results have been partially reported in \citep{tavani20,Tavani2021}. We monitored R3 in the X-rays (0.3–10 keV) daily during 15 activity cycles (12 of them falling within the time intervals of this work). XRT observations were carried out in Windowed Timing (WT) readout mode, with 2-7 daily pointings. The time resolution of WT data is 1.8 ms and each pointing has a typical exposure of $\sim$\,1.8 ks. We combined all the data within each proposal and processed them using the XRTDAS software package (v.3.7.0) \footnote{developed by the ASI Space Science Data Center (SSDC)} within the HEASoft package (v.6.30.1). The data were calibrated and cleaned with standard filtering criteria using the xrtpipeline task and the calibration files available from the Swift/XRT CALDB (version 20220803). The imaging analysis was executed selecting events in the energy channels between 0.3 and 10 keV and within a 20 pixel ($\sim47''$) radius, including the 90\% of the point-spread function.
The background was estimated from a nearby source-free circular region with the same radius value.

\section{Results} \label{sec:results}

\subsection{Radio} \label{subsec:resultradio}

\begin{table*}
\caption{Properties of the detected bursts. The second column reports the  barycentric time of arrival (TOA) at infinite frequency (TDB units, DE405 ephemeris, TAI clock, DM-delay constant $2.41 \times 10^{-16}$\,\pcm\,s, reference frequencies for the DM correction are $375.875$\,MHz for SRT-P and $500$\,MHz for uGMRT). The TOAs have been computed via the {\sf Tempo2} software \citep{tempo2}. The third column contains the bursts phase $\phi$ (see Sec.\,\ref{subsec:chromaticity}). The fourth column reports the FWHM width $\Delta t$ of the bursts. $S_{\nu}$, $F_{\nu}$, $E$ and $\dot{\nu}$ are respectively the peak flux density, the fluence density and the isotropic energy of the bursts. The last column reports the linear drift rate $\dot{\nu}$ of the bursts (for the bursts for which it was not possible to assess this value, i. e. a drift was not present, a N.A. value is reported).} 
\label{tab:burstsproperties}
\centering                          
\begin{tabular}{l c c c  c c c c}        
\hline\hline                 
Burst ID & TOA & $\phi$ & $\Delta t$  & $S_{\nu}$ & $F_{\nu}$ & $E$ & $\dot{\nu}$ \\
         & $\mathrm{[MJD]}$ & & $\mathrm{[ms]}$  & $\mathrm{[Jy]}$ & 
         $\mathrm{[Jy\,ms]}$ & $\mathrm{[10^{37}  erg]}$ & $\mathrm{[MHz \ ms^{-1}]}$ \\
\hline                        
SRT-P-01 & 59162.8253243218 & 0.5870 & 19 $\pm$ 2       &  3.1 $\pm$ 0.8 & 48 $\pm$ 16 & 8 $\pm$ 2.5 & N.A. \\ 
SRT-P-02 & 59162.8343320911 & 0.5875 & 14.0 $\pm$ 0.5   & 12.2 $\pm$ 3.0 & 139 $\pm$ 43 & 24 $\pm$ 7  & -9.7 $\pm$ 0.6  \\ 
SRT-P-03 & 59162.8567440946 & 0.5889 & 13.2 $\pm$ 0.5   & 11.7 $\pm$ 2.9 & 125 $\pm$ 39 & 21 $\pm$ 6 &  -11.4 $\pm$ 0.6  \\ 
SRT-P-04 & 59162.8665428914 & 0.5895 & 23 $\pm$ 2       & 3.2 $\pm$ 0.8 & 61 $\pm$ 20 & 10 $\pm$ 2  & N.A.  \\ 
SRT-P-05 & 59162.8752448474 & 0.5900 & 18 $\pm$ 2       & 5.3 $\pm$ 1.5 & 77 $\pm$ 29 & 13 $\pm$ 3  & N.A.  \\ 
         & 59162.8752449374 & 0.5900 & 23 $\pm$ 1       & 5.3 $\pm$ 1.5 & 98 $\pm$ 27 & 17 $\pm$ 3   &  N.A.  \\ 
SRT-P-06 & 59162.9481096844 & 0.5945 & 14 $\pm$ 2      & 5.5 $\pm$ 1.4 & 62 $\pm$ 23 & 10 $\pm$ 6 &  N.A.  \\ 
SRT-P-07 & 59162.9513861530 & 0.5947 & 23 $\pm$ 4       & 2.7 $\pm$ 0.7 & 50 $\pm$ 20 & 8 $\pm$ 1 &  N.A.  \\ 
SRT-P-08 & 59162.9515615864 & 0.5947 & 20 $\pm$ 2       & 3.0 $\pm$ 0.7 & 49 $\pm$ 17 & 8 $\pm$  1 &  N.A. \\ 
SRT-P-09 & 59162.9661486339 & 0.5956 & 12 $\pm$ 2      & 2.3 $\pm$ 0.6 & 22 $\pm$ 9 & 4 $\pm$ 3  & N.A.  \\ 
SRT-P-10 & 59194.7961833830 & 0.5448 &  9 $\pm$ 2       & 2.1 $\pm$ 0.6 & 15 $\pm$ 7 & 3 $\pm$ 1  & N.A.  \\ 
SRT-P-11 & 59244.9407380909 & 0.6155 & 22 $\pm$ 2       & 5.2 $\pm$ 1.3  & 92 $\pm$ 21 & 16 $\pm$ 2  & -11.3 $\pm$ 0.6 \\ 
SRT-P-12 & 59244.9503715983 & 0.6160 & 18.1 $\pm$ 0.9   & 11.8 $\pm$ 2.9 & 172 $\pm$ 55 & 29 $\pm$ 6  & -10.7 $\pm$ 0.6  \\ 
SRT-P-13 & 59244.9559004303 & 0.6164 & 12 $\pm$ 1       & 7.3 $\pm$ 1.9 & 70 $\pm$ 27 & 12 $\pm$ 3 & N.A.  \\ 
SRT-P-14 & 59258.8293625230 & 0.4660 & 16 $\pm$ 2       & 4.9 $\pm$ 1.3 & 64 $\pm$ 23 & 10 $\pm$ 2 & N.A.  \\ 
uGMRT-01 & 59441.0257263218 & 0.6231 & 20 $\pm$ 2      & 0.8 $\pm$ 0.2 & 8.9 $\pm$ 0.8 & 2.1 $\pm$ 0.2 & -5.9 $\pm$ 0.5  \\ 
uGMRT-02 & 59441.0453208005 & 0.6243 &  9.2 $\pm$ 0.7  & 0.6 $\pm$ 0.2 & 2.8 $\pm$ 0.9 & 0.3 $\pm$ 0.1 & N.A.  \\ 
uGMRT-03 & 59441.0575323455 & 0.6250 & 20.8 $\pm$ 0.6  & 7.9 $\pm$ 0.2 & 84 $\pm$ 3 & 36 $\pm$ 1 & -10.5 $\pm$ 0.5 \\ 
uGMRT-04 & 59441.0622128892 & 0.6254 &  7.6 $\pm$ 0.8  & 0.4 $\pm$ 0.4 & 2.5 $\pm$ 0.7 & 0.6 $\pm$ 0.1 & N.A.  \\ 
uGMRT-05 & 59441.0665624511 & 0.6256 &  6 $\pm$ 1      & 0.4 $\pm$ 0.1 & 0.6 $\pm$ 0.1 & 0.06 $\pm$ 0.01 & N.A. \\ 
uGMRT-06 & 59441.0944839544 & 0.6273 & 16 $\pm$ 4      & 1.3 $\pm$ 1.2 & 7.9 $\pm$ 2.3 & 0.8 $\pm$ 0.2 & N.A.   \\
uGMRT-07 & 59441.1011067833 & 0.6277 & 15 $\pm$ 8      & 0.2 $\pm$ 0.1 & 0.9 $\pm$ 0.8 & 0.2 $\pm$ 0.2 & N.A.   \\ 
\hline                               
\end{tabular}
\end{table*}

\subsubsection{Burst Properties} 
\label{subsec:burstproperties}

Throughout the whole campaign the radio instruments, SRT and uGMRT, detected a total of 21 bursts (14 from the SRT at P-band and seven with the uGMRT). Figures \ref{fig:srtbursts} and \ref{fig:gmrtbursts} show the dedispersed ($\rm DM = 348.82$\,\pcm) waterfall plots of the bursts. Their properties are reported in Tab.\,\ref{tab:burstsproperties}.



The widths of the bursts were computed as the full width at half maximum of a Gaussian function. In the case of the SRT, the data have been converted into flux density units applying the standard radiometer equation \citep{handbookpulsar}. In the case of uGMRT, as we mentioned in Sec. \ref{subsec:obsradio}, we calibrated the FRB data by multiplying it with a conversion factor counts-Jy estimated via on-source and off-source observations of 3C48 \citep{perley17}. 
The fluence densities $F_{\nu}$ were computed by integrating the calibrated background subtracted light curves within their respective burst widths.

The bursts detected by both radio telescopes do not possess complex structure, apart from the of SRT-P-05 and uGMRT-03 in which two sub-bursts are present, with the latter not clearly resolvable. Six of the 21 bursts show a downward drifting toward lower frequencies, as already seen from this source \citep{chawla20, chamma_r3_drift}. We evaluated the linear drift rate $\dot{\nu}$ by using a standard auto-correlation analysis \citep{hessels19}, obtaining values which scatter around an average value of $-10 \pm 1$ MHz ms$^{-1}$ for the four bursts of SRT-P and $-8.2 \pm 0.8$ MHz ms$^{-1}$ for the two bursts from uGMRT.

\subsubsection{L-band Observations and Frequency-Dependent Activity of R3} 
\label{subsec:chromaticity}

Regarding the SRT L-band, in addition to the standard single-pulse search, we carefully inspected the data for clustered bright pixels located $\sim$ 9.9 s (the DM-delay between the top-frequencies of SRT-L and SRT-P) before the bursts at P-band. We do not report occurrence of a burst simultaneous at two different radio frequencies, as already noticed for R3 by \cite{sand22_multiband,pilia20}. This is similar to what was observed for R1 \citep{majid_2020_r1dsn,caleb_2020_r1meerkat} although, interestingly, R1 was multi-band simultaneously detected once \citep{law_2017_r1multitel}.
For the whole campaign, by exploiting the radiometer equation, we set a fluence density UL (for a S/N=6, 1 ms burst) of about  0.4 Jy ms at L-band. At P-band, for the dates in which we had no detection, we set a fluence density UL of about 4.6 Jy ms. 

\begin{figure}
    \centering
    \includegraphics[width=\linewidth]{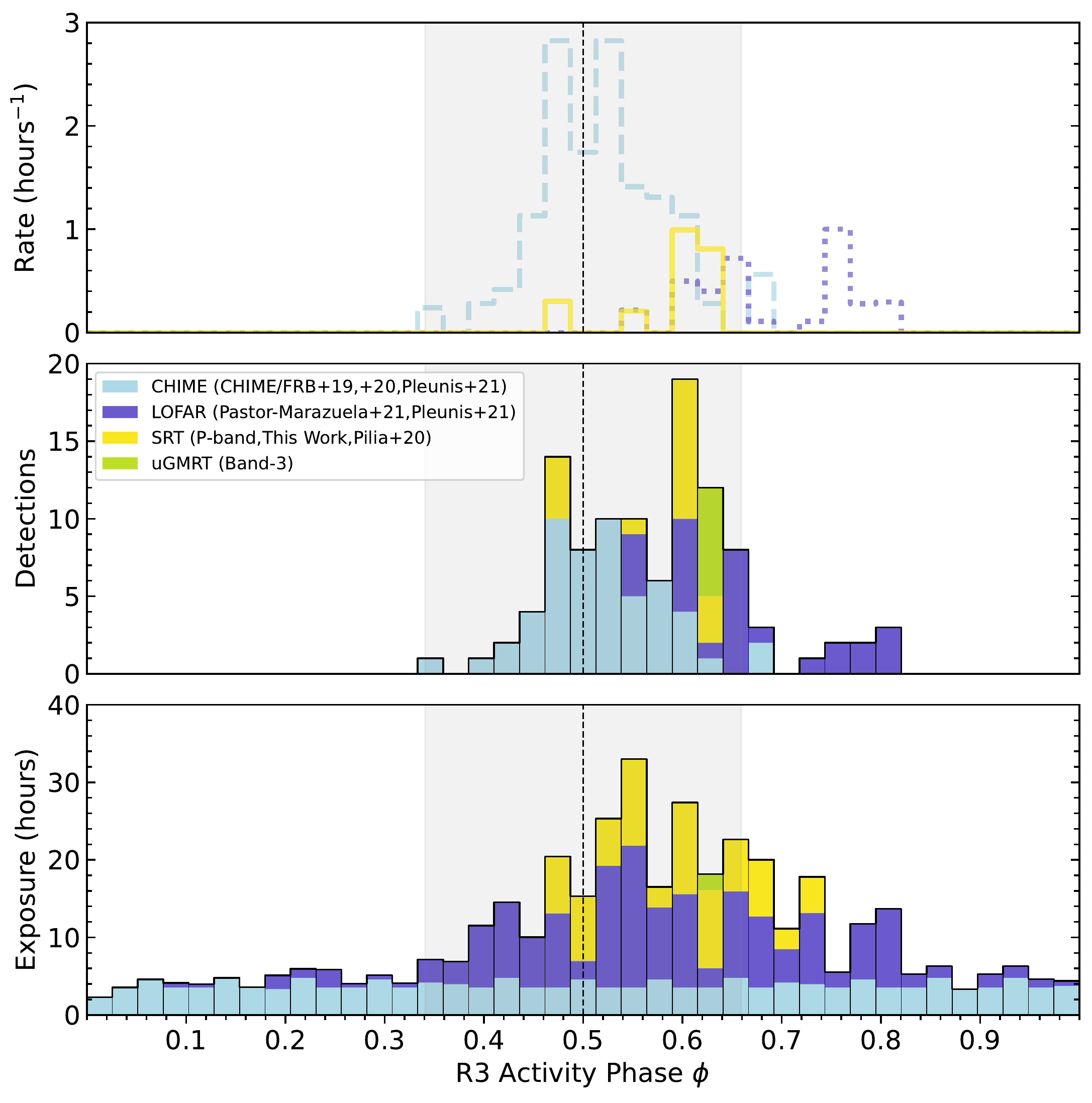}
    \caption{Number of burst detections (mid panel), exposure time (bottom panel) and events rate (top panel) as a function of the activity phase of R3 both for the SRT and the uGMRT. Data from CHIME \citep{pleunis21a,chime_2020_periodr3,chime_2019_8newfrb} and LOFAR \citep{pleunis21a,pastor-marazuela_2021_r3} are reported as comparison. The grey area depicts the CHIME predicted activity window.}
    \label{fig:phaseactivity}
\end{figure}

\begin{figure*}
    \centering
    \includegraphics[width=\linewidth]{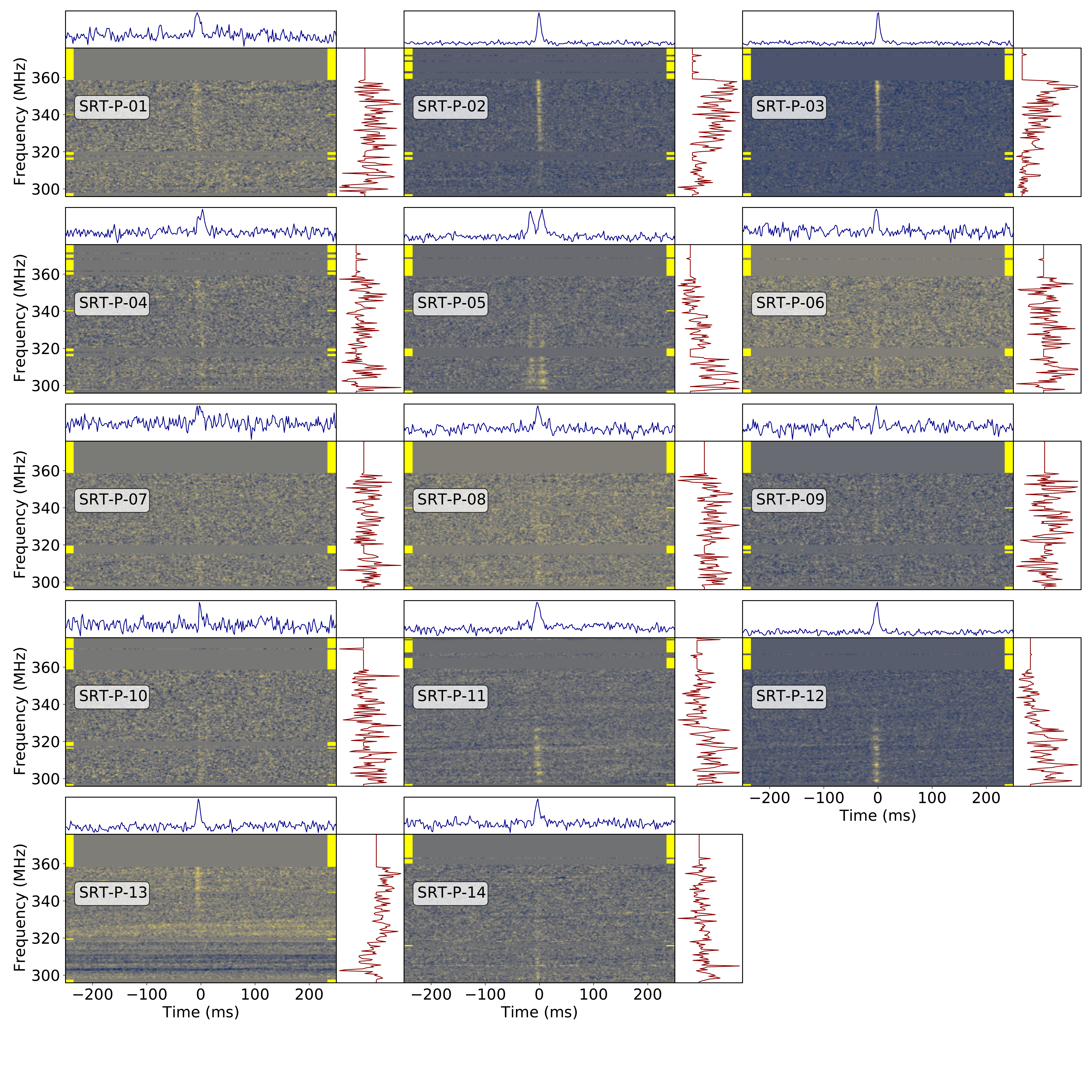}
    \caption{Bursts detected with the SRT. The data have a resolution of 0.5 MHz in frequency and 1.9 ms in time. For each burst, the top panel shows the frequency-averaged time series. The central panel is the spectrogram of the signal and the right panel is the time averaged (around the width of the burst) spectrum. The rows with the yellow ticks are masked channels due to RFI. }
    \label{fig:srtbursts}
\end{figure*}

\begin{figure*}
    \centering
    \includegraphics[width=\linewidth]{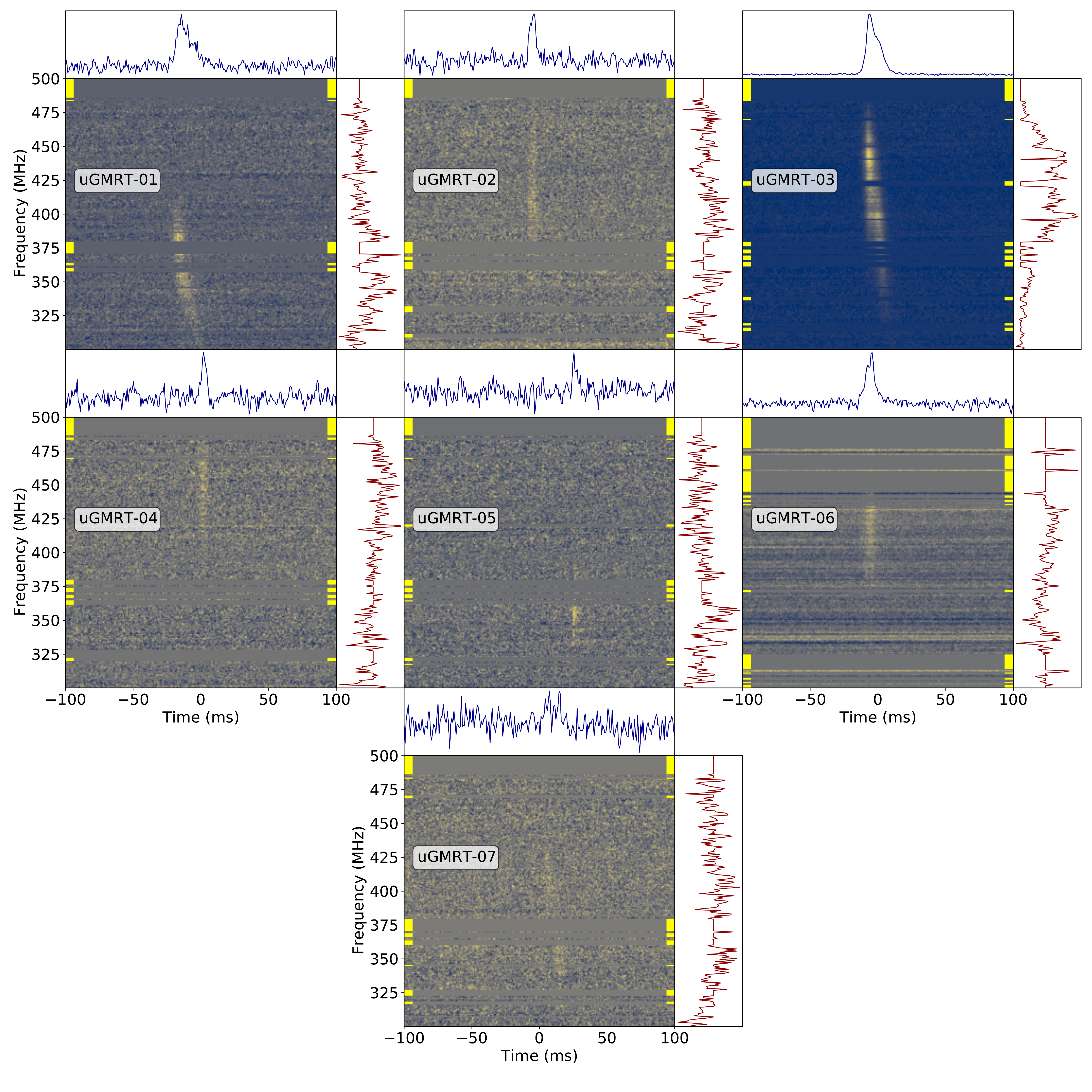}
    \caption{Bursts detected with the uGMRT. The data have a resolution of 0.8 MHz in frequency and 0.8 ms in time. For each burst, the top panel shows the frequency-averaged time series. The central panel is the spectrogram of the signal and the right panel is the time averaged (around the width of the burst) spectrum. The rows with the yellow ticks are masked channels due to RFI. }
    \label{fig:gmrtbursts}
\end{figure*}

Table \ref{tab:burstsproperties} shows the bursts phases ($\phi$), obtained by folding the time of arrivals (TOAs) at $P=16.33$ days and reference epoch of $58369.40$ MJD \citep{pleunis21a}. The SRT bursts at $336$ MHz have phases within the range of $0.46<\phi<0.62$, while all the uGMRT bursts have $\phi \sim 0.62$ as they were detected on the same day. Figure \ref{fig:phaseactivity} shows the number of bursts detected and the telescope exposure time as a function of the activity phase $\phi$. Since we only observed with the uGMRT for a single day, we only discuss the SRT case. In addition to our burst sample, we include the three bursts published by \cite{pilia20}, which were observed with the SRT using the same dual-band receiver (for a total observing time of $\sim 30$ hours) and detected at P-band. The SRT observations span a phase range $0.46 < \phi < 0.72$, with a total observing time of $69$\,hours distributed for $\sim 40$\% in $0.46<\phi<0.56$, $\sim 37$\% in $0.56<\phi<0.66$, and $\sim 23$\% in $\phi>0.66$. 70\% of our detections lie in the second phase range, with the rest being in the first one, which suggests that our low-frequency detections follow the frequency-dependent activity manifested by R3 (see Sec.~\ref{sec:introduction}). 

We compare our detections with the ones reported by CHIME in \cite{pleunis21a} (54 detections) at $600$ MHz and with the LOFAR bursts \citep[26 bursts,][]{pleunis21a, pastor-marazuela_2021_r3} at $150$ MHz. From the top panel of Fig.~\ref{fig:phaseactivity}, which depicts the number of detections normalized by the exposure time (i.e., the rate), we see that the SRT-P distribution appears compatible with both CHIME and LOFAR. To investigate whether our bursts are drawn from the same distributions as the one of the CHIME or LOFAR bursts, we performed a two-sample Anderson-Darling test \citep[AD, see e.g.][]{andersondarling}. By applying the AD test on the SRT-P/CHIME (SRT-P/LOFAR) dataset, we obtain a confidence level of only $2.8 \sigma$ ($1.6 \sigma$) indicating that the SRT-P bursts may not be drawn from the same distribution as CHIME (LOFAR). Based on these results, we conclude that our burst distribution, given the current dataset, is compatible with both CHIME and LOFAR, and further detections are necessary.

Our non-detections at L-band are consistent with the chromatic activity of R3, given the covered phase ranges, even though detections have been recorded at phases up to $\phi=0.5$  \citep{pastor-marazuela_2021_r3}.
Furthermore, our observations at L-band were performed with a channel width of $1$\,MHz, which retains an intra-channel smearing of $\sim 0.7$\,ms. 

Shorter bursts would have potentially been lost. 
Our non-detections are, again, consistent with the observed properties of R3 bursts, given that the bursts detected at higher frequencies \citep[$>1$\,GHz,][]{bethapudi22,pastor-marazuela_2021_r3} have  smaller temporal widths ($\lesssim 1$\,ms), comparable with our smearing, with respect to those at frequencies of a few hundred MHz, which have widths of the order of tens of milliseconds \citep{pastor-marazuela_2021_r3}. 


\subsubsection{Source Activity Rate}
\label{subsec:r3rate}

Of the 14 SRT detections, nine happened on the same day. We observed R3 for 5.8 hours on 2020/11/09, implying an average burst rate of $\sim$ 1.55 hours$^{-1}$ at the SRT-P frequencies and above a fluence density of $4.6$\,Jy\,ms. This value seems particularly high if compared, for instance, to the 2021/02/13 observation when we observed the source for 2 hours and detected only one event (corresponding to a burst rate of 0.5 hours$^{-1}$). In order to assess if the 2020/11/09 was a highly active day, as it frequently happens for R1 \citep{zhang_2018_r1ml, li_2021_1650r1,hewitt_2021_r1storm,jahns_2022_r1november}, we follow arguments similar to those discussed by \cite{trudu2022a} to obtain the rate of events expected by the SRT-P. We scale the  rate reported for the source by CHIME/FRB of $R_0 = 0.9 \pm 0.5$ hours$^{-1}$ at 600 MHz and above a fluence density of 5.2 Jy ms \citep{chime_2020_periodr3} to the SRT-P central frequency and fluence density limit. We consider  a value of $\beta =  1.5$ \citep{macquart19} as the spectral index and a slope $\alpha = 2.3$  \citep{chime_2020_periodr3} for the cumulative fluence distribution $N(>F) \propto F^{-\alpha + 1}$ obtaining: 
\begin{equation}
    \label{eq:rate}
    R(> F_{\nu}) = R_0 \times \zeta({\rm BW}, F_{\nu}^{\rm min}) 
    \times  \left( \frac{\nu_c}{600 \ {\rm MHz}} \right)^{- \beta}
    \times  \left( \frac{F_{\nu}}{5.2 \ {\rm Jy \ ms}} \right)^{-\alpha + 1 } \ . 
\end{equation}
Here we additionally introduced a coefficient $\zeta ({\rm BW}, F_{\nu}^{\rm min})$ to take into account the observational biases of the bandwidth BW of SRT-P to detect off-band events above its fluence density threshold $F_{\nu}^{\rm min}$ as discussed by \cite{aggarwal_2021_bandlimited}. We estimated $\zeta ({\rm BW}, F_{\nu}^{\rm min})$  via a Monte Carlo simulation. We model a burst from R3 as a Gaussian function in the frequency domain with the amplitude (energy), the frequency centroid and the FWHM width as free parameters.  We assume the energy distributed in the range $(10^{36}-10^{40})$\,erg as a negative power-law with the same index $\alpha$. The frequency centroids were drawn from a uniform distribution $\mathcal{U}(300,1000)$\,MHz. We excluded the LOFAR frequencies and frequencies $>$\,GHz to avoid to add to the model the frequency-dependent activity of the source (see Sec.\,\ref{subsec:chromaticity}). Lastly the FWHM frequency widths were assumed Normal distributed $\mathcal{N}(\mu,\sigma)$, with mean value $\mu = 107$\,MHz and  standard deviation $\sigma = 59$\,MHz. We obtained them considering the reported R3 frequency widths by CHIME/FRB \citep{mckinven22,pleunis21a,chime_2020_periodr3,chime_2019_8newfrb}: with its $400$\,MHz observational bandwidth, it provides the least biased frequency width measurements. We generated 1000 bursts and evaluated $\zeta$ as the fraction of these generated bursts which are detectable by the telescope (being hence above the telescope fluence density threshold), and repeated this for 1000 trials. By doing this, we obtain $\zeta (80 \ {\rm MHz}, 4.6 \ {\rm Jy\,ms}) = 0.33 \pm 0.04$ for the SRT-P.

Including all these parameters, Eq.\,\ref{eq:rate} yields a burst rate  $R(> {\rm 4.6 \ Jy\,ms}) = 0.7 \pm 0.4$ hours$^{-1}$ for SRT-P, implying that the 2020/11/09 and 2021/02/13 rates were not outliers.

\subsection{Optical}
\subsubsection{Asiago}\label{Asiagoresults}

We performed a search for any significant increase in the count rate on the 1 ms binned optical light curves of all acquisitions and, in particular,  around the time of the detected radio bursts. All nine bursts detected with SRT in 2020/11/09, the single burst detected in 2021/02/13 and the seven bursts detected with uGMRT in 2021/08/14 fall inside Aqueye+ and/or IFI+Iqueye observing windows. The acquisitions of 2021/08/14 were contaminated by the activity of the Perseid meteor shower.

To assess the significance of a peak, we follow the procedure adopted by \cite{zampieri22} to search for optical bursts at the time of the occurrence of hard X-ray burst from the magnetar SGR\,J1935+2154. We assume a Poisson distribution with the average rate of the observation and fix a 3$\sigma$ detection threshold $n_{\rm t,obs}$ corresponding to a chance probability of 0.0027/$N_{\rm trials}$ in any of the bins of the observation. In case a radio burst is detected, we calculate also a detection threshold $n_{\rm t}$ corresponding to a chance probability of 0.0027/$N_{\rm trials}$ in any of the bins during an interval of $\pm$100 ms and $\pm$15 s around the time of arrival of the burst. 
These two values were chosen to estimate the significance in the two different scenarios of an optical burst almost coincident with the radio burst, and a delayed one.
$N_{\rm trials}$ is the total number of bins in the interval. We obtain typical values of $n_{\rm t,obs}$ in the range 18--20 counts bin$^{-1}$ for Aqueye+ and 12--14 counts bin$^{-1}$ for IFI+Iqueye. For the observations in which radio bursts were detected, we obtain $n_{\rm t}(100 {\rm ms}) = $ 11--13 counts bin$^{-1}$ and $n_{\rm t}(15 {\rm s}) =$
15--16 counts bin$^{-1}$ for Aqueye+ (2020/11/09 and 2021/08/14; see Tab. \ref{tab:asiago}) and $n_{\rm t}(100 {\rm ms}) = $ 6--8 counts bin$^{-1}$ and 
$n_{\rm t}(15 {\rm s}) = $ 10--11  counts bin$^{-1}$ for IFI+Iqueye (2020/11/09 and 2021/02/13, see Tab. \ref{tab:asiago}).

\begin{figure}
    \centering
    \includegraphics[width=\linewidth]{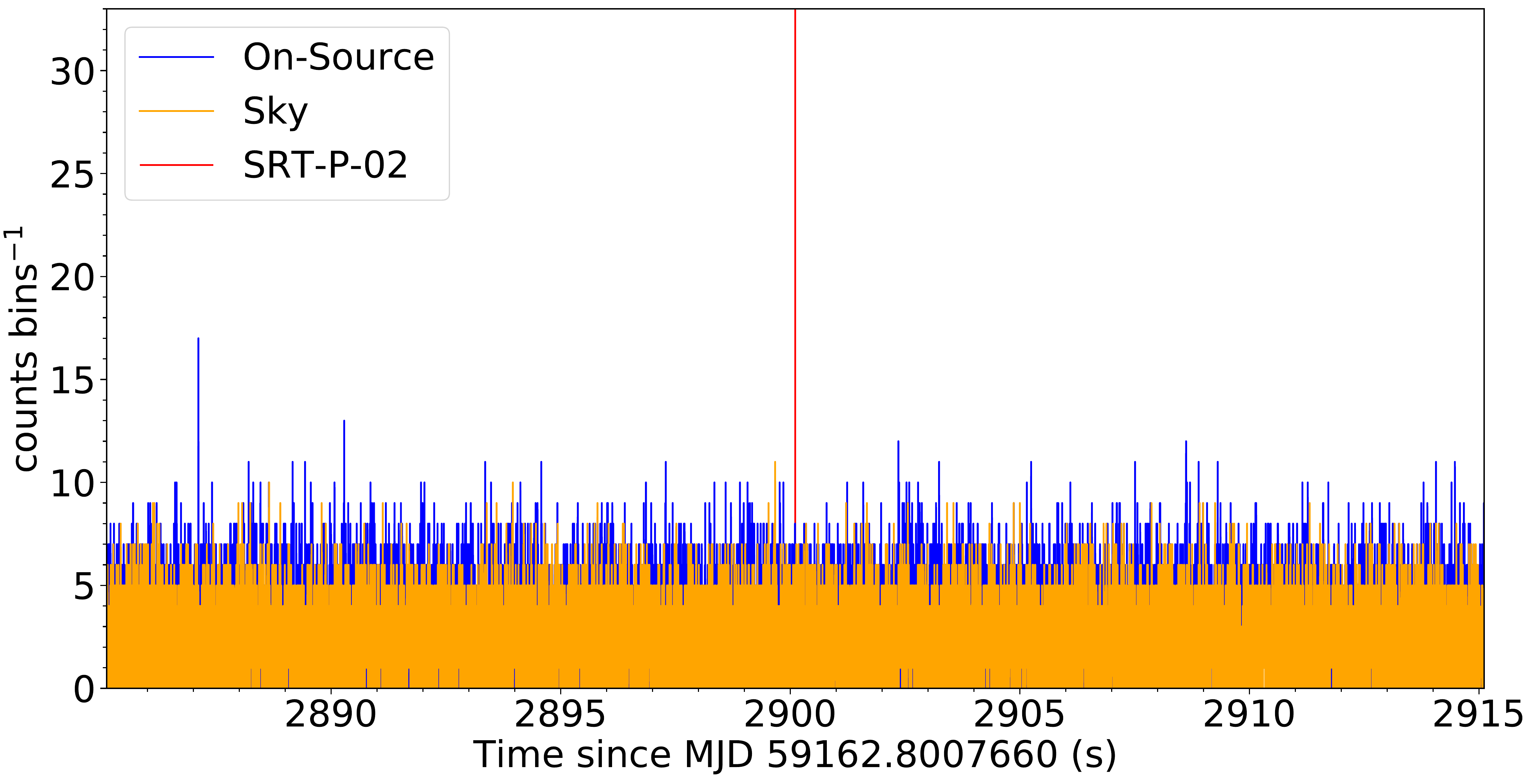}
    \caption{Light curves (binned at 1 ms) of the on-source (blue) and sky (orange) detectors of the Aqueye+ observation corresponding to R3, taken on 2020/11/09 (Obs ID 20201109-200635 in Tab. \ref{tab:asiago},the Aqueye+ starting time has been barycentred). An interval of $\pm$15 s around the time of arrival of the burst SRT-P-02 (red vertical line) is shown.}
    \label{fig1}
\end{figure}

Only one on-source optical peak with $n=18$ counts bin$^{-1}$ in an Aqueye+ observation falls inside an interval of $\pm 15$ s around the SRT-P-02 TOA. In barycentred time units, the optical leads the radio one by $13.971$\,s (see Fig.\,\ref{fig1}). This peak is slightly above threshold for the $\pm$15 s window ($n_{\rm t} = 15$ counts bin$^{-1}$); corresponding to a potential detection at the 90\% confidence level of an optical burst with 14.4 mag per ms, a fluence density of $0.007$\,Jy\,ms, and a luminosity of $1.6 \times 10^{43}$ erg s$^{-1}$), but slightly below threshold considering the entire observation ($n_{\rm t,obs} = 19$ counts bin$^{-1}$). In fact, two other peaks are detected in the same observation, each with $n=17$ counts bin$^{-1}$. We thus conclude that this optical flash is marginally significant, but not at a high enough level to consider it a robust counterpart of the radio burst.

To estimate an UL to the optical brightness during a radio burst, we consider the highest on-source-only peak occurring in an interval of $\pm$ 100 ms and $\pm$ 15 s around the time at which a burst is detected, after subtracting a rate per bin that has a high Poissonian probability (99.73\%) to be exceeded by chance in a single bin. This procedure returns an UL of 7--10 counts bin$^{-1}$ (in $\pm 100$ ms) and 12--16 counts bin$^{-1}$ (in $\pm 15$ s)  for the two Aqueye+ observations of 2020/11/09 and 7--9 counts bin$^{-1}$ (in $\pm 100$ ms) and 11--14 (in $\pm 15$ s) for the Aqueye+ observation of 2021/08/14. Using the $V$ band calibration of Aqueye+ \citep{zampieri16}, these values correspond to an average (non extinction corrected) optical brightness of $V_{\pm 100 {\rm ms}} 14.25-14.64$ mag per ms and $V_{\pm 15 {\rm s}} = 13.74-14.05$ mag per ms for 2020/11/09 and $V_{\pm 100 {\rm ms}} = 14.49-14.64$ mag per ms and $V_{\pm 15 {\rm s}} = 13.88-14.15$ mag per ms for 2021/08/14. The corresponding deepest non extinction-corrected upper limits to the fluence density are: $4.6 \times 10^{-15}$ erg~cm$^{-2}$ (0.005 Jy ms) for an interval of $\pm 100$ ms and $7.9 \times 10^{-15}$ erg~cm$^{-2}$ (0.009 Jy ms) for an interval of $\pm$15 s in November 2020; $4.6 \times 10^{-15}$ erg~cm$^{-2}$ (0.005 Jy ms) for an interval of $\pm$100 ms
and $7.2 \times 10^{-15}$ erg~cm$^{-2}$ (0.008 Jy ms) for an interval of $\pm 15$ s in August 2021. Assuming a distance of 149 Mpc \citep{marcote20}, these values eventually imply the following limits to the luminosity: $1.2 \times 10^{43}$ erg s$^{-1}$ for an interval of $\pm 100$ ms and $2.1 \times 10^{43}$ erg s$^{-1}$ for an interval of $\pm 15$ s on 2020/11/09; $1.2 \times 10^{43}$ erg~s$^{-1}$ for an interval of $\pm 100$ ms and $1.9 \times 10^{43}$ erg~s$^{-1}$ for an interval of $\pm 15$ s on 2021/08/14.

The lack of detection of any optical peak during the partially simultaneous observations of 2020/11/09 with IFI+Iqueye confirms these results, although the inferred ULs are less deep. IFI+Iqueye was the only active optical photometer in 2021/02/13, at the time of the SRT detection of SRT-P-14 (obs ID 20210213-195451; see Tab. \ref{tab:asiago}). The UL inferred for this observation are 7 counts bin$^{-1}$ in $\pm$ 100 ms and 10 counts bin$^{-1}$ in $\pm$ 15 s that, using the $V$ band calibration of IFI+Iqueye \citep{zampieri16} and the overall optical transmission efficiency at the time of the observation (32\%), corresponds to an average (non extinction corrected) optical brightness of $V_{\pm 100 \rm{ms}} = 11.6$ mag $V_{\pm 15 \rm{s}} = 11.2$ mag. The corresponding ULs to the fluence density are $7.9 \times 10^{-14}$ \,erg\,cm$^{-2}$ (0.09 Jy ms) and $1.1 \times 10^{-13}$ \,erg\,cm$^{-2}$ (0.13 Jy ms), respectively.
On the other hand, significant optical peaks of duration between $\sim$1 ms and $\sim$10 ms are detected during some observations. While some are simultaneously observed with the on-source and sky detectors and are therefore diffuse foreground atmospheric events (most likely meteors), others are on-source-only or sky-only events the nature of which is still under investigation. Results of this analysis will be reported elsewhere.

\subsubsection{CAHA 2.2m}



The $\simeq 2.4$ million 5-ms frames collected by CAHA 2.2m/AstraLux on 2021/01/13 were initially investigated for events deviating by more than $5\sigma$ from the background.
We performed aperture photometry at the FRB position after removing frames affected by multi-pixels cosmic rays, which were not automatically filtered out by the analysis software. A conservative $6\sigma$ level cut allowed us to identify a total of 28 candidate events.

To quantify the statistical significance of this value, we performed the same analysis on the thirty void regions shown in Fig.\,\ref{fig:AstraLux_regions}. The resulting number of candidates was similar.

\begin{figure}
    \centering
    \includegraphics[width=\linewidth]{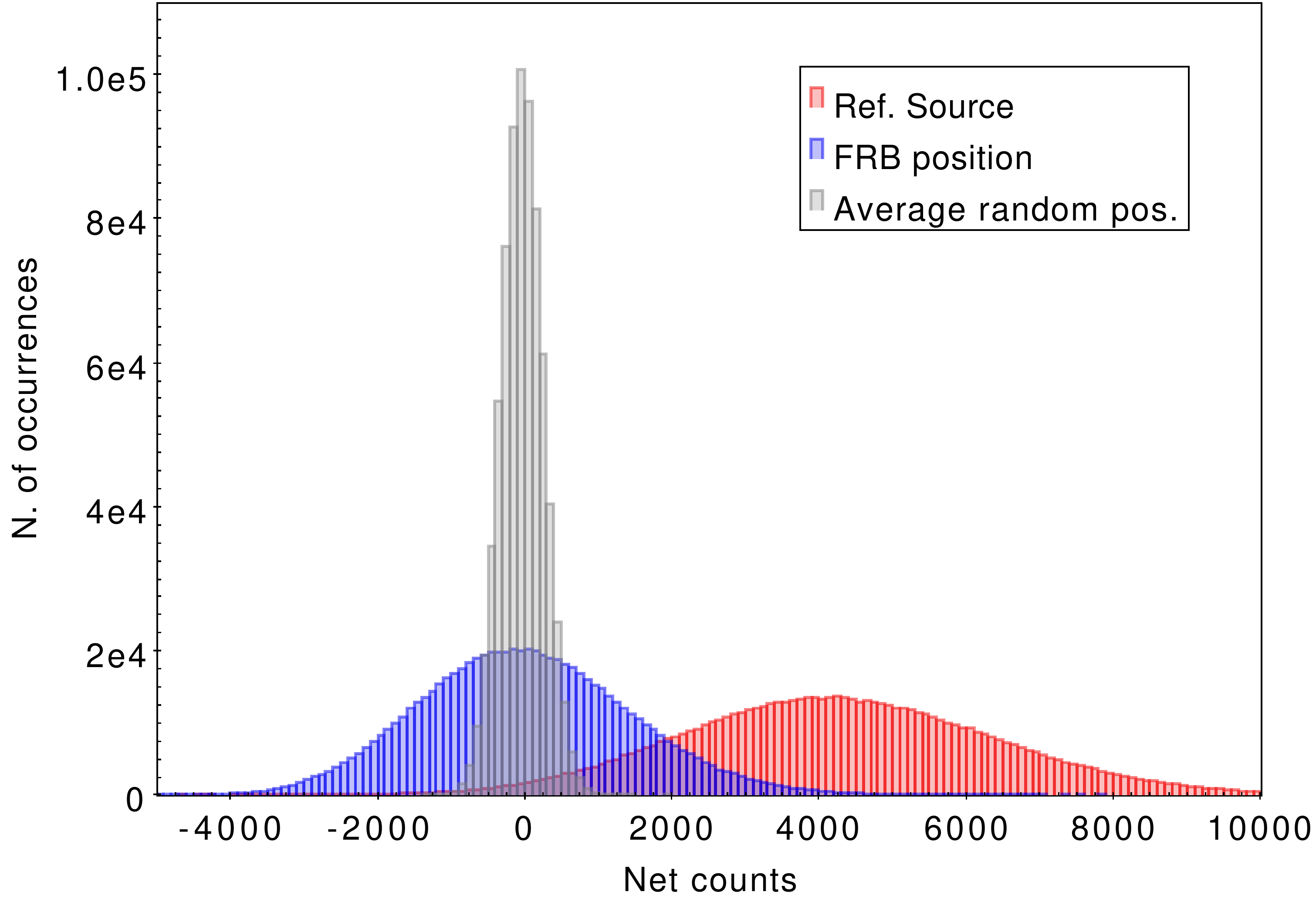}
    \caption{Count statistics comparison at the FRB (blue) and reference source (red) position. The grey histogram is obtained averaging the counts from the thirty random regions reported in Fig. \ref{fig:AstraLux_regions}. Here plotted are the initial 720,000 net counts from the 5-ms frames collected by AstraLux.}
    \label{fig:AstraLux_histo}
\end{figure}

As a cross-check, the light curves obtained summing the counts collected on the FRB, void and reference source regions were statistically investigated. As expected, the distribution of net counts (centre $\simeq 4200$, $\sigma = 2130$) and the corresponding $\sigma$ for the G $\simeq 15.6$ reference source clearly show its detection (see Fig.\,\ref{fig:AstraLux_histo}). In fact, the 1-s integrated light curve shows no sign of variability. On the other hand, for the FRB and void regions the bias subtracted count distributions are zero centred and narrower ($\sigma \sim 1400$). However, we note that only $\sim$ 1--2 values exceed the $6\sigma$ level ($\sim$ 10--15 above $5\sigma$). This is quite different from the figures obtained by the aperture photometry.
Visual inspection of the frames with $\sigma > 6$ at the FRB location revealed two cases that we can consider of interest, occurring at 19:47:29.7 and 22:29:50.9 UTC. However, lacking a simultaneous monitoring with any other instrument, we cannot claim a burst detection.
Based on the overall statistics of the reference source, we quote a limiting magnitude (White filter, not corrected for extinction) for the single 5 ms frame of 16.0 at $3\sigma$ and 15.5 at $5\sigma$. The latter corresponds to an upper limit on the putative source luminosity of $5.3 \times 10^{42}$ erg\,s$^{-1}$.

\subsubsection{CMO SAI MSU and RTT-150}
Over two nights we have accumulated about 12 million frames on the 2.5 m CMO telescope. For each frame we estimated the signal in a $2.5''$ (radius) aperture around the FRB position and also the background level in several selected regions. The mean signal in aperture was 20 electrons. Assuming a Poisson distribution
for the signal with this mean value, the standard deviation is 4.4 electrons. The probability to have more than one frame with signal larger than 48 electron from background in series of $1.2\times10^7$ frames is 0.37. We performed a search for flux excursions with this threshold on individual frames. After a manual screening of the selected frames (which were caused by cosmic rays and detector anomalies) no viable candidate bursts was left. Assuming a linear response of the camera we could place a conservative UL on optical emission from the FRB position at a level of 16.5 G magnitude per 4 ms frame. Taking into account the fact that the normalised response curve of our instrument deviates from the Gaia G band response curve by no more than $11\%$, we can put constraints on mean flux density: $8.9\times10^{-16}$~erg\,s$^{-1}$\,cm$^{-2}$\,\AA$^{-1}$. 

A similar analysis to the one described above was performed over the RTT-150 data. Apart from a few large amplitude outliers caused by the cosmic rays or CCD glitches no viable candidate events were found, providing an 5$\sigma$ UL of about 15.5 G magnitude. 




\subsubsection{TNG}
Seven bursts were detected by the uGMRT during the observation carried out on the night of 2021/08/14 (see Tab.\,\ref{tab:burstsproperties}) during which SiFAP2/TNG was observing for about 70 \% of the total radio exposure time. As in the case of Asiago simultaneous observation (see Sec.\,\ref{Asiagoresults} for more details), also our data were affected by the Perseid meteor shower, causing several high-significance spikes to be randomly distributed within the dataset of SiFAP2. We detected them on both detectors of SiFAP2 monitoring the target and nearby sky background, respectively (see Sec.\,\ref{TNG}). To search for possible optical counterparts of radio bursts, time of arrivals of single optical (320--900\,nm) photons were rebinned at 1\,ms resolution and computing our ULs with this resolution to be consistent with the results obtained by Asiago.


We searched for the optical counterpart in two different time intervals ($\pm 15$\,s and $\pm 100$\,ms) around the epoch of the detection of all the seven bursts detected by uGMRT. Since no significant peaks were found at the time of bursts, we put only an upper limit on the magnitude. This value was computed for each burst by adopting the same procedure as that reported in Sec.\,\ref{Asiagoresults}.
Following the SiFAP2 calibration curve\footnote{https://www.tng.iac.es/instruments/sifap2/}, we obtained a non dereddened UL magnitude of V = 15.96, and V = 16.42, for the $\pm 15$\,s and $\pm 100$\,ms time intervals, respectively. The corresponding fluence densities are 1.56\,mJy\,ms and 1.02\,mJy\,ms. 



\subsection{High-Energy} \label{subsec:highenergyresult}

\subsubsection{\emph{AGILE}}
\label{subsec:agileresult}

We searched in archival Super-A, MCAL and GRID data, at the times reported in Table \ref{tab:burstsproperties}, to look for coincident X-ray and $\gamma$-ray emission from the source.  As we reported in Table \ref{tab:AGILEcov}, \emph{AGILE} has partial coverage of the 21 bursts reported. Of the 12 bursts in the MCAL FoV, we found no evidence of significant signals in the data. We could provide standard MCAL fluence ULs at 3$\sigma$ when a data interval is present, and sub-ms fluence ULs when no data acquisition is available. Moreover, due to the AGILE spinning and to the reduced Super-A and GRID FoV with respect to the MCAL FoV, only 1 event falls inside the GRID exposed region and none in the Super-A one. However this burst falls within a data acquisition gap and hence we cannot provide a punctual UL. We could provide also GRID 3$\sigma$ ULs in ${\rm E} \geq 100$\,MeV on two integrations: on 6 days, obtaining UL$_{\rm 6days} = 6.0 \times 10^{-11}$\,erg\,cm$^{-2}$\,s$^{-1}$, and on about 17 months from 2020/04/01 to 2021/09/15, with UL$_{\rm 17months} = 7.7 \times 10^{-12}$\,erg\,cm$^{-2}$\,s$^{-1}$.

\subsubsection{\emph{Insight--\/}HXMT}
\label{subsec:HXMTresult}
We carried out two kinds of search: (i) a blind one, regardless of the radio bursts, by applying the criteria that have already been tailored and applied to \emph{Insight--\/}HXMT data in previous investigations \citep{Guidorzi2020}; (ii) a targeted search around the times of the FRBs that were covered with Insight-HXMT observations. The expected background counts were estimated following the prescriptions of \citet{Guidorzi2020}.
We did not find any statistically significant ($>3\sigma$ confidence) candidate in either case. We hereafter focus on the results of (ii), whereas a more detailed report on the results of (i) will be reported elsewhere.

We observed simultaneously with \emph{Insight--\/}HXMT 7 out of the 21 FRBs reported in this work (Table~\ref{tab:burstsproperties}). In particular, we covered simultaneously with both LE and ME 5 FRBs detected with the SRT (SRT-P-01, SRT-P-02, SRT-P-07, SRT-P-08, and SRT-P-09), whereas uGMRT-01 was covered with all of the three \emph{Insight--\/}HXMT instruments and uGMRT-02 with the ME only.
For each of these 7 FRBs we carried out a search for statistically significant excesses within a time window centred on the FRB time with a duration of 200~s or shorter, depending on the availability of data. To account for the unknown duration of any possible high-energy counterpart, we spanned a logarithmically-spaced range of bin times, from 1, 2, 4,\ldots to 128 ms. In each case we applied a threshold on the counts of each individual detector as well as to the summed counts, by imposing a detection significance of $p_{\rm sing}/N_{\rm bins}$, with $p_{\rm sing}$ as close as possible to $2.7\times10^{-3}$ (i.e., $3 \sigma$ confidence), taking into account the multitrial due to the number of bins $N_{\rm bins}$ to be screened. Since the expected background counts in each instrument are $\ll 1$ over a few ms and because of the discreteness of Poisson distribution, the probability associated with a given threshold on the number of counts varies enormously even by changing the threshold by one: this is why the corresponding significance can only approximately be $3 \sigma$ for most cases.

In no case we found counts in excess of the thresholds. We therefore calculated the threshold counts on the summed light curves (LE+ME for the 5 SRT FRBs, LE+ME+HE for uGMRT-01, and ME only for uGMRT-02) for three representative bin times (1, 16, and 128 ms). We then converted these ULs on counts to corresponding fluence values by considering three different spectral models that were discussed by \citet{Guidorzi2020} and that can be plausibly expected for sources like magnetars: a power-law ({\sc pl}) with photon index $\Gamma=2$, an optically thin thermal bremsstrahlung ({\sc ottb}) with either $kT=50$ keV or $kT=200$ keV.

Table \ref{tab:Xupplim} reports the corresponding fluence limits in either $1-30$, or $1-100$, or $10-30$ keV energy passbands, depending on whether data from LE and ME only were available, or from all of the three instruments, or from the ME only, respectively. Overall, for each of the three integration times mentioned above, the corresponding $\approx 3\sigma$ ULs on the $1-30$ keV fluence are in the range $(1.6-2.9) \times 10^{-10}$ erg cm$^{-2}$ (1 ms), $(2.7-5.5) \times 10^{-10}$ erg cm$^{-2}$ (16 ms), and $(5.5-12) \times 10^{-10}$ erg cm$^{-2}$ (128 ms).

\begin{figure}
    \centering
    \includegraphics[width=\linewidth]{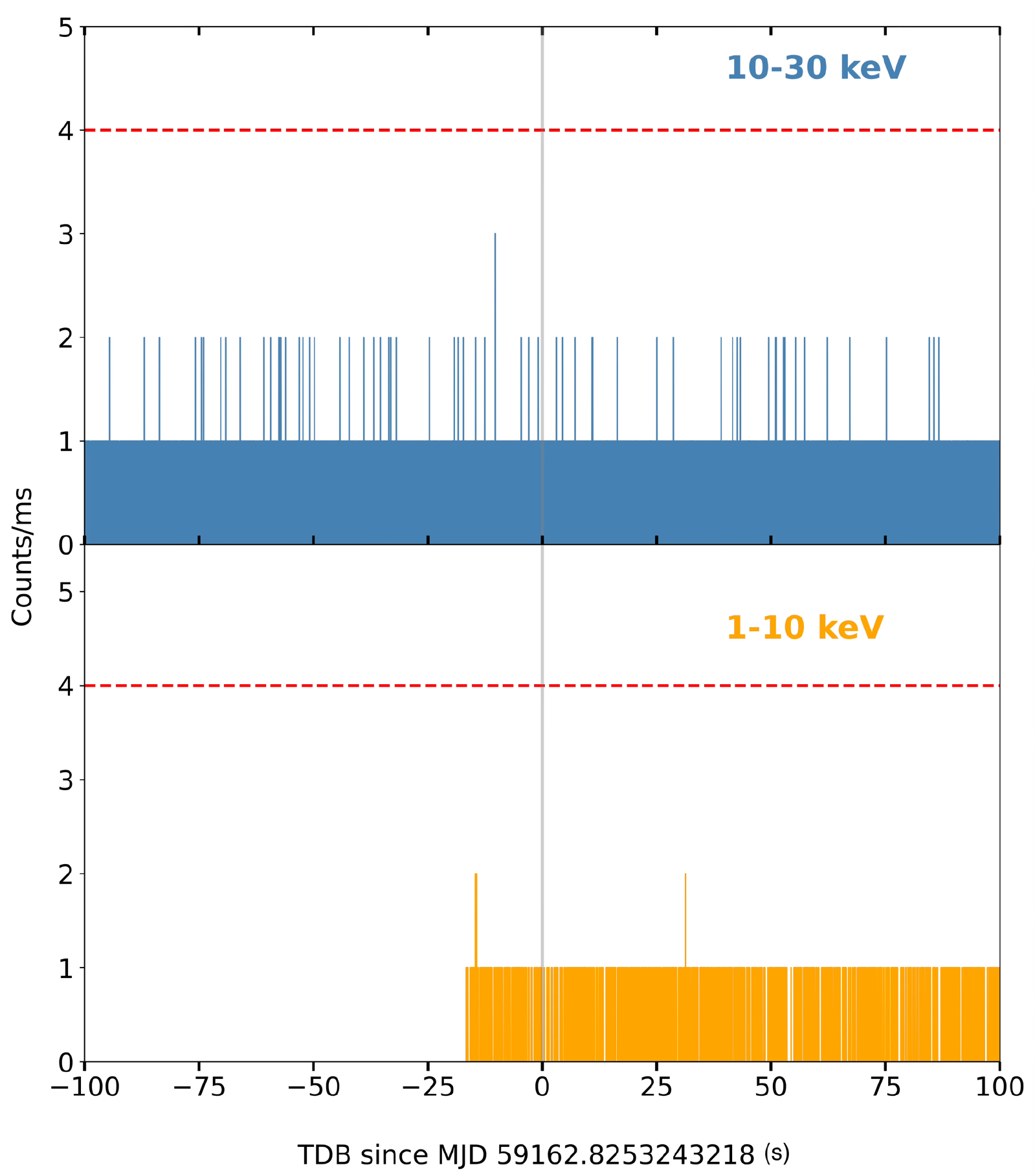}
    \caption{\emph{Insight--\/}HXMT 1-ms light curves of LE and ME obtained around SRT-P-01 (grey line). Dashed lines show the $\approx 3\sigma$ upper limits that account for the multi-trial search.}
    \label{fig:FRB1_HXMT}
\end{figure}

Figure \ref{fig:FRB1_HXMT} shows the \emph{Insight--\/}HXMT coverage with 1 ms resolution of SRT-P-01 with LE and ME, along with the corresponding $\approx 3\sigma$ detection threshold. 

\subsubsection{INTEGRAL}
\label{subsec:INTEGRALresult}

In the complete integrated mosaics of the INTEGRAL Soft Gamma-ray Imager (ISGRI) and the Joint European X-ray Monitor (JEM-X) we do not detect bursts from R3 and we use ISGRI observations to set an UL of 1.1$\times$10$^{-11}$~erg~cm$^{-2}$~s$^{-1}$ on average source flux in 28~--~80~keV band assuming a power-law spectrum with a slope of $2$. The limit on average source flux with JEM-X can be set in 3~--~30~keV band at the level 3.6$\times$10$^{-11}$~erg~cm$^{-2}$~s$^{-1}$.

In order to search for possible variable emission, we built ISGRI light curves on timescales of 1000 seconds and 1 second, in 28-80 keV energy range. We do not detect any variable emission on these timescales and put ULs of 2.7$\times$10$^{-10}$~erg~cm$^{-2}$~s$^{-1}$ and 1.1$\times$10$^{-8}$~erg~cm$^{-2}$~s$^{-1}$ on 1000 seconds and 1 second long impulsive emission.

Unfortunately, none of the bursts occurred when R3 was in the FoV of INTEGRAL pointing instruments. However, INTEGRAL all-sky instruments were taking data during all of the radio events, and we derive an UL on 1-second long hard X-ray burst within 10~seconds from each radio burst, as shown in Tab.~\ref{tab:integralul}. 

\subsubsection{Swift}
\label{subsec:swiftresult}

We carried out a search for the presence of an X-ray source candidate at the R3 position in the whole Swift/XRT WT mode dataset. No X-ray source was detected (a $>3\sigma$ confidence was required). We therefore extracted  the 3\,$\sigma$ countrate ULs using the XIMAGE package (sosta command) and converted to fluxes using a standard single power-law spectral model with photon index 2.0, and correcting for absorption for a column density N$_H$ fixed to the Galactic value of 7.1\,$\times\,10^{21} \rm cm^{-2}$ \citep{HI4PI_2016} redshifted for the known z = 0.0337 \citep{marcote20}. The X-ray observations exposure and the corresponding ULs are reported in Tab.~\ref{tab:swiftcov}. 

\section{Discussion}
\label{sec:discussion}

\label{sub:mwldiscussion} 

\begin{figure}
    \centering
    \includegraphics[width=\linewidth]{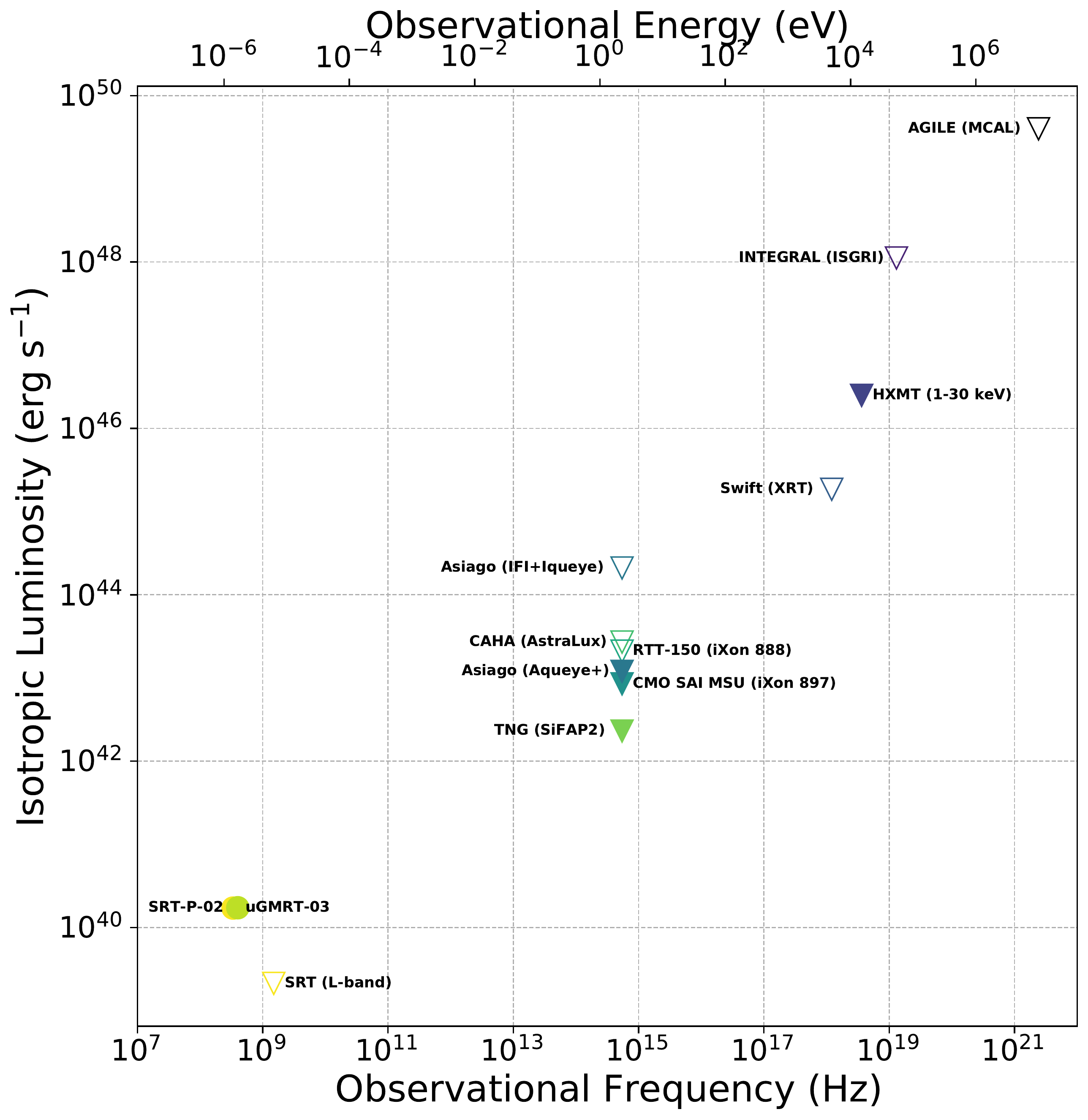}
    \caption{Burst isotropic luminosity as function of the observational frequency for all the instruments involved in the MWL campaign. The dots are the luminosities of the bursts SRT-P-02 and uGMRT-03. The empty triangle markers represent the non punctual upper limits (ULs), that is coincident with no radio burts. The filled triangles are punctual ULs, coincident with a radio burst detection (see Sec.\,\ref{sub:mwldiscussion}). All optical ULs are converted in the V band and for a 1 ms burst.}
    \label{fig:uls}
\end{figure}

Figure \ref{fig:uls} shows an overview of the isotropic luminosity ULs by all the instruments involved in this MWL campaign. We can use these punctual ULs to constrain the optical/radio and high-energy/radio energy ratios.  SRT-P-02 was observed simultaneously with Aqueye+ and CMO SAI MSU. Assuming that the burst width is of the same order of tens of ms in both bands, this implies a ratio $\xi =  E_{\text{optical}} / E_{\text{radio}}< 7.8 \times 10^2$ and $\xi  < 5.6 \times 10^2$, respectively. CMO SAI MSU UL however, is obtained from a camera (iXon 897) while Asiago uses a fast photometer (Aqueye+) and the observations are performed and reduced via different techniques. In the case of the search for an optical counterpart of a signal like an FRB, due to its short duration, a fast photometer is more capable to explore such low duration ranges.    
The radio burst uGMRT-03 was simultaneous with Aqueye+ and SiFAP2 observations and, with their punctual UL, we can constrain the ratio $\xi < 6.9 \times 10^2$ for Aqueye+ and $\xi < 1.3 \times 10^2$ for SiFAP2. SiFAP2 UL on R3 is the most stringent UL on the optical emission for this source to date.
The previous most stringent UL in the optical band for R3 was from \cite{andreoni_2020_r3optical}. They performed untargeted observations of R3, not simultaneous with a radio instrument, with the Zwicky Transient Facility, monitoring the source in active windows and providing an UL of $\xi < 10^8$. 

\cite{beloborodov_2020} describes a scenario where  young, hyperactive magnetars are the progenitors of FRBs. Their magnetic flares are able to generate blast waves in the surrounding wind medium and this could result into optical flashes, of the duration of up to $\sim 1$~s, simultaneous to the radio bursts. Given their formation, in the tail of a preceding flare, they are expected during recurrent flaring episodes, and they are expected to be weak unless they are formed after a strong explosion, when they could reach $E_{\text{optical}}\sim 10^{44}$\,erg. Our current UL, for both dates when more bursts were observed, at the moment can only exclude such strong flares. 


SRT-P-02 was also simultaneous with \emph{Insight--\/}HXMT observations, which provided a punctual UL (1--30 keV band) on the released isotropic energy $E_{\rm X-ray}$,  in the range of $(2-3) \times 10^{45}$\,erg for a time bin of $128$\,ms, depending on which emission model is considered (see Sec.\ref{subsec:HXMTresult}). This UL is of the same order of magnitude of the ones placed by Chandra \citep{scholz20} and XMM-Newton \citep{pilia20} for the same source also simultaneously with a radio burst. In the  case of the SRT-P-02 burst, this UL implies an UL in the X-ray/radio efficiency $\eta = E_{\text{X}} / E_{\text{radio}} < (0.9-1.3) \times 10^7$.

It is worth to make a comparison with the case of the SGR\,J1935+2154 event. With its bi-chromatic simultaneous detection in radio \citep{chime20sgr, bochenek20} and at high-energies  \citep{mereghetti20,Tavani2021,hxmt_sgr,ridnaia20}, it provided evidence that magnetars can be FRB emitters and hence be the sources behind at least some of the FRBs observed. In that case, the energy released for the burst detected by CHIME was $\sim 3 \times 10^{34}$ erg and it corresponded, in the 1--250 keV band of \emph{Insight--\/}HXMT, to $\sim 1 \times 10^{39}$ erg \citep{hxmt_sgr}, implying $\eta \sim 10^5$. This value is two orders of magnitude smaller than the $\eta$ UL that we reached with our MWL observations for R3. With this current UL we cannot exclude X-ray/radio efficiencies $\eta$ ten times greater or of the same order of magnitude of the efficiency of the SGR\,J1935+2154 event.

If we compare the results of our multi-cycle MWL campaign on R3 to the evidence based on the single event associated to the SGR\,J1935+2154 burst, we conclude that the detection of a R3 radio burst with fluence $ \gtrsim 10^3$\,Jy\,ms would allow us to set constraints on $\eta$ matching those of the Galactic event, in light of the sensitivity thresholds of current X-ray telescopes. 
If we determine the threshold of a given X-ray instrument as $F_{\rm X-ray}^*$, we can express the corresponding radio fluence density $F_{\nu}$, with the assumption of an efficiency $\eta$ as:
\begin{equation}
    \label{eq:fxtoradio} 
    F_{\nu} = \frac{F_{\rm X-ray}^*}{\eta \Delta \nu} \ ,
\end{equation}
where $\Delta \nu$ is the frequency width of the radio burst. We assume the average value of $\Delta \nu = 107 \pm 59$\,MHz as discussed in Sec.\ref{subsec:resultradio}. We can now compute via Eq.\ref{eq:rate} the rate of events $R(>F_{\nu})$ with radio fluence density greater than $F_{\nu}$ for the SRT-P and uGMRT Band 3. By performing the same Monte Carlo simulation for uGMRT Band 3, as discussed in Sec.\,\ref{subsec:r3rate}, we obtain a value of $\zeta (200 \ {\rm MHz}, 0.6 \ {\rm Jy\,ms}) = 0.43 \pm 0.04$. 

\begin{figure*}
    \centering
    \includegraphics[width=\linewidth]{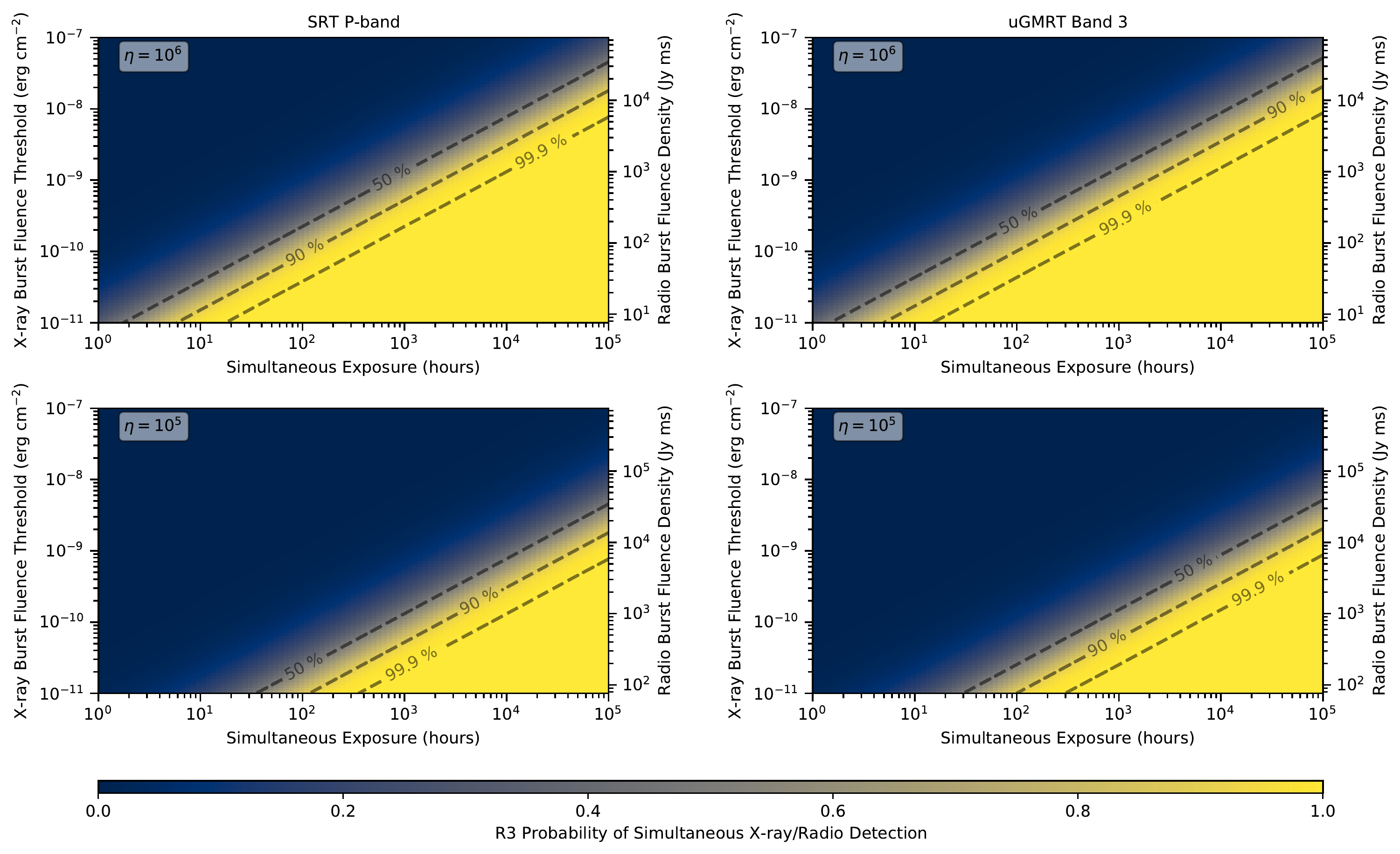}
    \caption{Probability of detecting a X-ray/radio event given a certain X-ray/radio efficiency $\eta$ observing in the radio band with the SRT-P (left) or the uGMRT Band 3 (right). Supposing that the event is above the fluence X-ray threshold (left y-axis) the colormap shows the probability of detecting a radio burst with radio fluence (right y-axis), constrained by the efficiency $\eta$, as a function of the simultaneous exposure (x-axis). The dashed lines represent the iso-probability contour plots for a probability of 50, 90 and 99.9 \%.}
    \label{fig:r3mwlprob}
\end{figure*}

Adopting the assumption that the R3 bursting activity is Poissonian, using the expected rate $R(>F_{\nu})$, we can roughly estimate the detection probability $p(>F_{\nu})$ of detecting one or more bursts with fluence density $>F_{\nu}$ in a campaign of duration $\Delta T$:
\begin{equation}
    \label{eq:probability} 
    p(>F_{\nu}) = 1 - e^{-R(>F_{\nu}) \Delta T} \ .
\end{equation}
Figure \ref{fig:r3mwlprob} shows the probability computed via Eq.\,\ref{eq:probability} for an X-ray detection simultaneous with the SRT-P or the uGMRT Band 3 as a function of the time of simultaneous exposure. Evaluating how much a certain X-ray satellite is sensitive to a putative X-ray burst requires several assumptions on the emission mechanism of the source in order to get a punctual value for $F_{\rm X-ray}^*$. Keeping the discussion to the level of orders of magnitude, the X-ray telescopes involved in the campaign were capable to detect bursts with a fluence threshold in the range of $10^{-10}-10^{-7}$\,erg\,cm$^{-2}$. Considering a 99.9 \% detection probability, this would require X-ray/radio campaigns $3 \times 10^2 - 3 \times 10^6$\,hours long for $\eta = 10^6$, or $6 \times 10^3 - 5 \times 10^7$\,hours for $\eta = 10^5$, with respect to the X-ray thresholds considered.

\section{Summary and Conclusions}
\label{sec:conclusion}

This work presents the results of a multi-wavelength observational campaign on FRB\,20180916B (R3) performed between October 2020 and August 2021. The campaign involved the SRT and the uGMRT radio telescopes, the Galileo and Copernico (Asiago), CMO SAI MSU, CAHA 2.2m and RTT-150 optical telescopes, and lastly the high-energy satellites AGILE, \emph{Insight--\/}HXMT, INTEGRAL and Swift. 

We detected 14 new bursts with the SRT at 336 MHz and seven new bursts with the uGMRT at 400 MHz. Neither the optical instruments nor the high-energy ones detected statically significant bursts at their frequencies. Asiago reports an optical peak within a window of $\pm15$\,s around the second radio burst detected by the SRT on the night of 2020/11/09. The peak is above the threshold of the aforementioned window of $30$\,s of 15 counts bin$^{-1}$. However, it is slightly below the 19 counts bin$^{-1}$ threshold obtained considering the statistics of the entire observation, thus not making it possible to robustly tag it as the optical counterpart of the radio burst.

We constrained the optical/radio efficiency $\xi$ setting a punctual upper limit (UL) thanks to a radio detection with the uGMRT and a simultaneous observation with the SiFAP2 photometer installed at the TNG, constraining $\xi < 1.3 \times 10^2 $, reporting the deepest UL in the optical band for this source.

\emph{Insight--\/}HXMT observations, simultaneous with a radio burst detected by the SRT, set a punctual UL on the X-ray/radio efficiency $\eta$  in the range of $\eta < (0.9-1.3) \times 10^7$ (1--30 keV band) depending on which emission model is used for the X-ray emission. Assuming the efficiency $\eta $ is either $10^6$ or $10^5$ (with the latter being comparable to that of the SGR\,J1935+2154 event), we can estimate what the required simultaneous X-rays/radio observing time and X-ray burst fluence should be to reach a given detection probability, with the instrumentation available in our campaign. Assuming that the events follow a Poissonian statistics, for X-ray sensitivities in the range of  $10^{-10}-10^{-7}$\,erg\,cm$^{-2}$, simultaneous X-ray/radio campaigns should last between $3 \times 10^2 - 3 \times 10^6$\,hours for $\eta = 10^6$, or $6 \times 10^3 - 5 \times 10^7$\,hours for $\eta = 10^5$. This result poses challenges in the detectability of a putative X-ray burst from this source. This can mainly be attributed to the cosmological nature of R3, being distant about $150$\,Mpc, $\sim 10^4$ times farther than the only source which to date has emitted an FRB-like signal simultaneous with an X-ray burst: SGR\,J1935+2154.
Despite the existence of closer-by repeaters (e.g. FRB\,20200120E, \citealt{kirsten_2022_20e_loc}; FRB\,20181030A, \citealt{bhardwaj2021b}), 
R3 represents at present a golden source for MWL follow-up for its periodic activity, high repetition rate and the relatively high energetic of the bursts.

\begin{acknowledgements}
The authors thanks the anonymous referee for the useful comments which significantly improved the quality of this work. M.T and M.P. wish to thank Paz Beniamini for useful discussions. M.T. acknowledges financial support from the PRIN MIUR 2017 Grant (MIUR Proj. 20179ZF5KS, CUP C24I19002290001) of the Italian Ministry for Universities and Research.
This work made use of the data from the \emph{Insight--\/}HXMT mission, a project funded by China National Space Administration (CNSA) and the Chinese Academy of Sciences (CAS). The \emph{Insight--\/}HXMT team gratefully acknowledges the support from the National Program on Key Research and Development Project (Grant No. 2016YFA0400800) from the Minister of Science and Technology of China (MOST) and the Strategic Priority Research Program of the Chinese Academy of Sciences (Grant No. XDB23040400). The authors thank supports from the National Natural Science Foundation of China under Grant Nos. 11503027, 11673023, 11733009, U1838201, and U1838202. The observations of R3 reported in the present paper were obtained under the approved program A03050096 within the \emph{Insight--\/}HXMT AO3 call and partly under the approved program A04050011 within the AO4 call (PI: C.~Guidorzi).
A.G. acknowledges financial support from the State Agency for Research of the Spanish MCIU through the “Center of Excellence Severo Ochoa” award for the Instituto de Astrofísica de Andalucía (SEV-2017-0709) and from national project PGC2018-095049-B-C21 (MCIU/AEI/FEDER, UE).
The observations made with the AstraLux instrument at the CAHA 2.2m telescope through the DDT.A20.267 were collected at the Centro Astronómico Hispano en Andalucía (CAHA) in Calar Alto,  jointly operated by the Instituto de Astrofísica de Andalucía (CSIC) and the Junta de Andalucía.
I.M. and A.L. acknowledge the financial support of the Russian Foundation of Basic Research (proj. 19-29-11029).
F.P. acknowledges financial support under the INTEGRAL ASI-INAF agreement 2019-35-HH.0 and ASI/INAF n. 2017-14-H.0.
F. A. and A.Pa. acknowledge financial support from the Italian Space Agency (ASI) and National Institute for Astrophysics (INAF) under agreements ASI-INAF I/037/12/0 and ASI-INAF n.2017-14-H.0, from INAF ’Sostegno alla ricerca scientifica main streams dell’INAF’, Presidential Decree 43/2018, and from the Italian Ministry of University and Research (MUR), PRIN 2020 (prot. 2020BRP57Z).

\end{acknowledgements}

\bibliographystyle{aa}
\bibliography{biblioMT}

%
%

\begin{appendix}

\section{Additional Instrumental Material: Observation Logs and Upper Limits}

\begin{table*}
\caption{Log of the optical observations of R3 carried out with Aqueye+@Copernicus and IFI+Iqueye@Galileo. The observations in bold are the ones in which a radio burst has been detected.}
\label{tab:asiago}
\centering                          
\begin{tabular}{l l c c}        
\hline\hline                 
Camera (Telescope) & Observation ID & Start Time (topocentric) & Exposure \\    
       &                & $\mathrm{[UTC]}$         & $\mathrm{[s]}$ \\    
\hline  
Aqueye+ (Copernicus) & 20201025-033103$_{-}$gti1 & 2020-10-25 03:01:54.000  &  1748 \\
 & 20201025-045249$_{-}$gti1 & 2020-10-25 03:52:51.000  &  520  \\
 &  {\bf 20201109-200635} & {\bf 2020-11-09 19:06:36.000}  &  {\bf 7198} \\
 &  20201109-221239$_{-}$gti1 & 2020-11-09 21:26:00.000  &  2797.000 \\
 &  {\bf 20201109-231359} & {\bf 2020-11-09 22:14:00.000}  &  {\bf 8997} \\
 &   20201110-020901 & 2020-11-10 01:09:02.000  &  1797 \\
 &  20201110-024652 & 2020-11-10 01:46:53.000  &  897.000  \\
 &  20201110-192232$_{-}$gti1 & 2020-11-10 19:28:24.000  &  6848 \\
 &  20201110-224746 & 2020-11-10 21:47:47.000  &  10798 \\
 &  20201111-014901 & 2020-11-11 00:49:03.000  &  1798 \\
 &  20201111-023825 & 2020-11-11 01:38:26.000  &  1198 \\
 &  20201111-231309$_{-}$gti1 & 2020-11-11 22:32:20.000 & 2050 \\
 &  20201112-005752$_{-}$gti1 & 2020-11-11 23:59:23.000 & 3508 \\
 &  20201112-015938 & 2020-11-12 00:59:40.000 & 3598 \\
 &  20201112-193512 & 2020-11-12 18:35:14.000 & 3598 \\
 &  20201112-205147 & 2020-11-12 19:51:48.000 & 3597 \\
 &  20201113-214903 & 2020-11-13 20:49:04.000 & 3597 \\
 &  20210111-193142 & 2021-01-11 18:31:44.000  &  7981 \\
 &  20210111-220228$_{-}$gti1 & 2021-01-11 21:02:29.000  &  1950 \\
 &  20210112-190452 & 2021-01-12 18:04:53.000  &  7197 \\
 & 20210112-210529 & 2021-01-12 20:05:31.000  &  7198 \\
 &  20210113-190245$_{-}$gti1 & 2021-01-13 18:02:46.000  &  850  \\
 &  20210113-210411$_{-}$gti1 & 2021-01-13 20:04:13.000  &  1650 \\
 &  20210113-210411$_{-}$gti2 & 2021-01-13 20:34:13.000  &  2300 \\
 &  20210114-183758$_{-}$gti1 & 2021-01-14 17:43:49.000  &  5300 \\
 &  20210813-232725 & 2021-08-13 21:27:26.000  &  7197 \\
 &  20210814-012913 & 2021-08-13 23:29:14.000  &  7197 \\
 &  20210814-234903 & 2021-08-14 21:49:04.000  &  7197 \\
 &  {\bf 20210815-014951} & {\bf 2021-08-14 23:49:53.000}  &  {\bf 7198} \\
 &  20210815-035032 & 2021-08-15 01:50:34.000  &  898  \\
 &  20210815-040612 & 2021-08-15 02:06:14.000  &  898  \\
 IFI+Iqueye (Galileo) &  {\bf 20201109-211141} & {\bf 59162 20:11:43.0}  & {\bf 7197} \\
 &  {\bf 20201109-233529$_{-}$gti1}  & {\bf 59162 22:35:31.0}  & {\bf 6850} \\
 &  20201110-013640               & 59163 00:36:42.0  & 2697 \\
 &  20201110-193505$_{-}$gti1     & 59163 18:35:07.0  & 7700 \\
 &  20201110-223616               & 59163 21:36:19.0  & 1797 \\
 &  20201110-233217               & 59163 22:32:19.0  & 3597 \\
 &  {\bf 20210213-195451}           & {\bf 59258 18:54:53.0}  & {\bf 5398} \\
 &  20210213-212539$_{-}$gti1     & 59258 20:47:21.0  & 2897 \\
 &  20210214-192834               & 59259 18:28:37.0  & 7197 \\
 &  20210214-212932               & 59259 20:29:34.0  & 4197 \\
 &  20210215-191649               & 59260 18:16:51.0  & 7197 \\
 &  20210215-211741               & 59260 20:17:44.0  & 5397 \\ 
\hline 
\end{tabular}
\end{table*}

\begin{table*}
\centering  
\caption{Observation log of CMO SAI MSU, RTT-150, CAHA 2.2m and TNG.}
\label{tab:optilog}
\begin{tabular}{l c c }        
\hline\hline                 
Camera (Telescope) & Start Time (topocentric)  & Exposure \\
       &                         & $\mathrm{[ks]}$ \\ 
\hline        
iXon 897 (CMO SAI MSU) & 2020-11-09 17:31:00.0 & 3.84 \\
                     & 2020-11-09 19:18:00.0 & 29.28 \\
                     & 2020-11-10 17:42:00.0 & 24.48 \\

iXon 888 (RTT-150)     & 2020-11-10 10:00:00.0 & 27.00 \\

AstraLux (CAHA 2.2m)   & 2021-01-13 19:24:25.0 & 3.60 \\
   & 2021-01-13 20:36:01.0 & 3.60 \\
   & 2021-01-13 21:47:36.0 & 3.60 \\
   & 2021-01-13 22:59:19.0 & 1.08 \\

SiFAP2 (TNG)           & 2021-08-15 00:35:00.0 & 8.7 \\
\hline 
\end{tabular}
\end{table*}

\begin{table*}
\caption[AGILE coverages and fluence upper limits]{\emph{AGILE} Coverage and MCAL upper limits (ULs) during the R3 radio bursts. The second, and fourth columns report the presence/absence of the source in the FoV of the onboard detector, respectively. The third column reports the presence/absence of MCAL data acquisition. ULs ($3\sigma$) are evaluated on existing MCAL data acquisitions at the burst times, and correspond to a fluence of 3$\sigma$ above the background; ULs (sub-ms) are evaluated on data acquisition if present, otherwise they refer to the UL fluences which would be required to issue a trigger with the onboard sub-ms MCAL trigger logic timescale \citep[see][]{2022ApJ...924...80U}. In presence of MCAL data acquisition, we show UL intervals obtained applying the spatial response matrix corresponding to source direction at burst times.}
\label{tab:AGILEcov}
\centering                          
\begin{tabular}{l c c c c c}        
\hline\hline                 
Burst & MCAL & MCAL & GRID  & UL & UL \\
        ID & FoV & D.A. & FoV &  ($3\sigma$) & (sub-ms) \\
         & & & & $\mathrm{[erg \ cm^{-2} ] }$ & $\mathrm{[erg \ cm^{-2} ] }$ \\
\hline
SRT-P-01 & NO & NO & NO    & \multicolumn{2}{c}{earth occultation}  \\
SRT-P-02 & NO & NO & NO    & \multicolumn{2}{c}{earth occultation}  \\
SRT-P-03 & YES & NO & NO   & --- & 1.15$\times10^{-8}$    \\
SRT-P-04 & YES & NO & YES  & --- & 2.55$\times10^{-8}$    \\
SRT-P-05 & YES & NO & NO   & --- & 6.88$\times10^{-8}$    \\
SRT-P-06 & NO & NO & NO    & \multicolumn{2}{c}{earth occultation}       \\
SRT-P-07 & NO & NO & NO    & \multicolumn{2}{c}{earth occultation}      \\
SRT-P-08 & NO & NO & NO    & \multicolumn{2}{c}{earth occultation}      \\
SRT-P-09 & YES & NO & NO   & --- & 6.88$\times10^{-8}$    \\
SRT-P-10 & YES & NO & NO   & --- & 6.88$\times10^{-8}$    \\
SRT-P-11 & YES & YES & NO   &  1.07$\times10^{-6}$ & 1.03$\times10^{-7}$ \\
SRT-P-12  & NO & NO & NO   & \multicolumn{2}{c}{idle mode}      \\     
SRT-P-13 & YES & NO & NO   & --- & 1.83$\times10^{-8}$    \\
SRT-P-14 & YES & NO & NO   & --- & 3.60$\times10^{-8}$    \\
uGMRT-01 & YES & YES & NO   &  1.03$\times10^{-6}$  & 1.00$\times10^{-7}$ \\
uGMRT-02 & NO & NO  & NO   & \multicolumn{2}{c}{earth occultation}  \\
uGMRT-03 & NO & NO & NO    & \multicolumn{2}{c}{earth occultation}  \\
uGMRT-04 & NO & NO & NO    & \multicolumn{2}{c}{idle mode}      \\   
uGMRT-05 & NO & NO & NO    & \multicolumn{2}{c}{idle mode}      \\   
uGMRT-06 & YES & YES & NO   & 1.09$\times10^{-7}$ & 1.06$\times10^{-8}$ \\
uGMRT-07 & YES & YES & NO   & 2.36$\times10^{-7}$ & 2.29$\times10^{-8}$ \\
\hline 
\end{tabular}
\end{table*}

\begin{table*}
\centering  
\caption{\emph{Insight--\/}HXMT observation log. }
\label{tab:Xobslog}
\begin{tabular}{l c c c c c}        
\hline\hline                 
Obs ID & Start Time (topocentric) & Stop Time (topocentric)  & LE Exposure & ME Exposure & HE Exposure \\
         &      &      & $\mathrm{[ks]}$ & $\mathrm{[ks]}$ & $\mathrm{[ks]}$ \\
\hline
P0303077001 & 2020-10-23 17:21:49 & 2020-10-24 01:29:27 & $11.1$ & $12.7$ & $7.9$\\
P0303077002 & 2020-10-24 17:12:47 & 2020-10-25 01:19:58 & $11.7$ & $12.3$ & $7.9$\\
P0303077003 & 2020-11-09 16:29:24 & 2020-11-10 00:36:49 & $14.1$ & $11.6$ & $4.4$\\
P0303077004 & 2020-11-10 16:21:02 & 2020-11-11 00:29:44 & $14.9$ & $12.1$ & $6.9$\\
P0303077005 & 2021-01-12 16:55:51 & 2021-01-13 01:12:32 & $11.2$ & $12.3$ & $10.3$\\
P0303077007 & 2021-01-13 16:46:54 & 2021-01-14 00:53:14 & $11.0$ & $12.3$ & $11.0$\\
P0403084001 & 2021-08-13 19:37:53 & 2021-08-15 02:10:55 & $53.5$ & $45.5$ & $31.9$\\
\hline 
\end{tabular}
\end{table*}

\begin{table*}
\centering
\caption{INTEGRAL observation log.}
\label{tab:integralobs}
\begin{tabular}{l c c c}        
\hline\hline                 
Start Time (topocentric) & Stop Time (topocentric) & Pointing ID & Exposure  \\
     &    &             & $\mathrm{[ks]}$ \\
\hline
                2020-12-18 21:42 & 2020-12-18 22:12 & 231000350010 -- 231000380010 & 7.2 \\
                2020-12-19 00:23 & 2020-12-19 00:53 & 231000400010 -- 231000400010 & 1.8 \\
                2020-12-19 01:27 & 2020-12-19 01:57 & 231000420010 -- 231000540010 & 23.3 \\
                2020-12-21 08:39 & 2020-12-21 09:09 & 231100250010 -- 231100280010 & 7.2 \\
                2020-12-21 11:38 & 2020-12-21 12:08 & 231100310010 -- 231100540010 & 43.1 \\
                2020-12-24 05:56 & 2020-12-24 06:26 & 231200350010 -- 231200380010 & 7.2 \\
                2020-12-24 08:36 & 2020-12-24 09:06 & 231200400010 -- 231200440010 & 9.0 \\
                2020-12-27 08:13 & 2020-12-27 08:44 & 231300540010 -- 231300560010 & 7.1 \\
                2020-12-27 10:52 & 2020-12-27 11:21 & 231300580010 -- 231300780010 & 37.7 \\
                2020-12-28 06:29 & 2020-12-28 06:59 & 231300950010 -- 231301010010 & 12.6 \\
                2020-12-30 13:29 & 2020-12-30 13:59 & 231400810010 -- 231400820010 & 3.6 \\
                2020-12-30 15:23 & 2020-12-30 15:53 & 231400850010 -- 231401040010 & 35.9 \\
                2021-01-01 02:33 & 2021-01-01 03:03 & 231500300010 -- 231500360010 & 12.6 \\
                2021-01-03 20:01 & 2021-01-03 20:31 & 231600330010 -- 231600360010 & 7.2 \\
                2021-01-03 22:42 & 2021-01-03 23:12 & 231600380010 -- 231600520010 & 27.0 \\
                2021-01-05 19:28 & 2021-01-05 19:59 & 231700020010 -- 231700060010 & 9.1 \\
                2021-01-12 06:55 & 2021-01-12 07:21 & 231900540010 -- 231900740010 & 70.9 \\
                2021-01-13 03:48 & 2021-01-13 04:45 & 231900750010 -- 231900800010 & 20.9 \\
                2021-01-15 00:01 & 2021-01-15 01:26 & 232000340010 -- 232000580010 & 88.4 \\
                2021-01-16 11:21 & 2021-01-16 11:51 & 232100030010 -- 232100200010 & 32.4 \\
                2021-01-16 22:03 & 2021-01-16 22:33 & 232100230010 -- 232100230010 & 1.8 \\
                2021-01-17 01:01 & 2021-01-17 01:31 & 232100290010 -- 232100300010 & 3.6 \\
                2021-01-22 17:44 & 2021-01-22 18:25 & 232300480010 -- 232300710010 & 61.4 \\
                2021-01-23 13:02 & 2021-01-23 13:32 & 232300750010 -- 232300800010 & 10.8 \\
                2021-01-23 18:22 & 2021-01-23 18:52 & 232300850010 -- 232300890010 & 9.0 \\
                2021-01-24 00:15 & 2021-01-24 00:45 & 232300960010 -- 232300960010 & 1.8 \\
                2021-02-14 17:03 & 2021-02-14 18:01 & 233200020010 -- 233200240010 & 84.3 \\
                2021-02-15 17:29 & 2021-02-15 18:27 & 233200260010 -- 233200300010 & 21.7 \\
\hline
\end{tabular}
\end{table*}

\begin{table*}
\centering
\caption{INTEGRAL Upper Limits (ULs). The second column is the ID of the Science Window Data catalogue, the third column contains the off-axis angles the fourth column contains the ULs for each burst, as discussed in \ref{subsubsec:INTEGRAL}.}
\label{tab:integralul}
\begin{tabular}{l c c c}        
\hline\hline                 
Burst ID & ScW ID & $(\theta,\varphi)$ & UL  \\
         &        & $\mathrm{[deg.]}$  & $\mathrm{[erg cm^{-2} s^{-1}]}$ \\
\hline

                SRT-P-01 & 229500410010 & 156.4, -159.4 &  4.8$\times$10$^{-7}$ \\
                SRT-P-02 & 229500410010 & 156.4, -159.4 &  4.2$\times$10$^{-7}$ \\
                SRT-P-03 & 229500420010 & 156.0, -155.3 &  5.2$\times$10$^{-7}$ \\
                SRT-P-04 & 229500420010 & 156.0, -155.3 &  4.6$\times$10$^{-7}$ \\
                SRT-P-05 & 229500420010 & 156.0, -155.3 &  5.3$\times$10$^{-7}$ \\
                SRT-P-06 & 229500440010 & 154.3, -160.2 &  4.9$\times$10$^{-7}$ \\
                SRT-P-07 & 229500440010 & 154.3, -160.2 &  4.1$\times$10$^{-7}$ \\
                SRT-P-08 & 229500440010 & 154.3, -160.2 &  5.6$\times$10$^{-7}$ \\
                SRT-P-09 & 229500440010 & 154.3, -160.2 &  3.8$\times$10$^{-7}$ \\
                SRT-P-10 & 230700380010 & 117.1, 147.8 &  4.3$\times$10$^{-7}$ \\
                SRT-P-11 & 232600520010 & 118.7, -118.3 &  5.5$\times$10$^{-7}$ \\
                SRT-P-12 & 232600530010 & 118.7, -118.3 &  4.8$\times$10$^{-7}$ \\
                SRT-P-13 & 232600530010 & 118.7, -118.3 &  3.7$\times$10$^{-7}$ \\
                SRT-P-14 & 233100630010 & 50.7, -70.8 &  1.3$\times$10$^{-7}$ \\
                uGMRT-01 & 240000130010 & 75.6, 93.6 &  2$\times$10$^{-7}$ \\
                uGMRT-02 & 240000130010 & 75.6, 93.6 &  1.1$\times$10$^{-7}$ \\
                uGMRT-03 & 240000140010 & 73.4, 94.2 &  1.8$\times$10$^{-7}$ \\
                uGMRT-04 & 240000140010 & 73.4, 94.2 &  1.5$\times$10$^{-7}$ \\
                uGMRT-05 & 240000140010 & 73.4, 94.2 &  1.4$\times$10$^{-7}$ \\
                uGMRT-06 & 240000150021 & 73.1, 95.0 &  2.4$\times$10$^{-7}$ \\
                uGMRT-07 & 240000150021 & 73.1, 95.0 &  1.5$\times$10$^{-7}$ \\
\hline 
\end{tabular}
\end{table*}

\begin{table*}
\centering
\caption{Swift Coverage and Flux Upper Limits (ULs, 0.3--10 keV band).}
\label{tab:swiftcov}
\begin{tabular}{c c c }        
\hline\hline                 
Start Time (topocentric) & Stop Time (topocentric) &  UL  \\
           &           &  $\mathrm{[erg\; cm^{-2}\, s^{-1}]}$ \\
\hline
2020-10-22 05:33:59 & 2020-10-27 18:09:56 & 3.0 $\times 10^{-13}$ \\
2020-11-08 17:16:50 & 2020-11-12 15:31:56 & 2.7 $\times 10^{-13}$ \\  
2020-12-09 04:06:58 & 2020-12-15 11:34:56 & 3.7 $\times 10^{-13}$ \\   
2020-12-30 18:17:55 & 2021-01-01 11:49:56 & 4.3 $\times 10^{-13}$ \\   
2021-01-10 02:40:33 & 2021-01-16 02:25:56 & 2.6 $\times 10^{-13}$ \\   
2021-01-29 16:25:31 & 2021-02-02 16:25:55 & 3.1 $\times 10^{-13}$ \\   
2021-03-02 13:27:42 & 2021-03-06 14:34:56 & 3.0 $\times 10^{-13}$ \\   
2021-03-17 13:16:26 & 2021-03-23 15:57:56 & 2.6 $\times 10^{-13}$ \\   
2021-04-19 12:02:25 & 2021-04-25 11:26:56 & 2.6 $\times 10^{-13}$ \\   
2021-05-09 09:21:59 & 2021-05-14 10:40:56 & 3.3 $\times 10^{-13}$ \\   
2021-05-24 09:11:17 & 2021-05-25 10:49:56 & 6.8 $\times 10^{-13}$ \\   
2021-06-07 08:07:52 & 2021-06-13 07:48:56 & 2.7 $\times 10^{-13}$ \\   
2021-06-22 04:47:04 & 2021-06-28 07:17:56 & 2.5 $\times 10^{-13}$ \\   
2021-07-08 06:05:06 & 2021-07-14 07:12:56 & 3.4 $\times 10^{-13}$ \\   
2021-07-25 04:47:26 & 2021-07-31 10:47:56 & 3.0 $\times 10^{-13}$ \\
\hline 
\end{tabular}
\end{table*}

\onecolumn{
\begin{longtable}{lccccc}
\caption{\emph{Insight--\/}HXMT high-energy upper limits (ULs) for the six FRBs that were observed.} \\ \hline \hline
\label{tab:Xupplim}
Burst ID & Spectral Model\tablefootmark{a} & Time Interval\tablefootmark{b} & Bin Time & Fluence UL (1-30 keV) & Confidence Level \\
         &      &   $\mathrm{[s]}$   &  $\mathrm{[ms]}$ & $\mathrm{[10^{-10} \ erg \ cm^{-2}]}$ & (Gaussian 2$\sigma$) \\
\hline
\endfirsthead
\caption{continued.} \\
\hline\hline
Burst ID & Spectral Model\tablefootmark{a} & Time Interval\tablefootmark{b} & Bin Time & Fluence UL (1--30 keV) & Confidence Level \\
         &      &   $\mathrm{[s]}$   &  $\mathrm{[ms]}$ & $\mathrm{[10^{-10} \ erg \ cm^{-2}]}$ & (Gaussian 2$\sigma$) \\
\hline
\endhead
\hline
\endfoot
\hline
\endlastfoot
 SRT-P-01 & {\sc pl} $\Gamma=2$ & $[ -16, +100]$ &  1  & $1.6$  & $2.8$\\         
 SRT-P-01 & {\sc pl} $\Gamma=2$ & $[ -16, +100]$ & 16  & $2.7$  & $2.5$\\
 SRT-P-01 & {\sc pl} $\Gamma=2$ & $[ -16, +100]$ & 128 & $6.6$  & $3.0$\\
 SRT-P-01 & {\sc ottb} $kT=50$~keV & $[ -16, +100]$ &  1  & $2.2$  & $2.8$\\         
 SRT-P-01 & {\sc ottb} $kT=50$~keV & $[ -16, +100]$ & 16  & $3.9$  & $2.5$\\
 SRT-P-01 & {\sc ottb} $kT=50$~keV & $[ -16, +100]$ & 128 & $9.5$  & $3.0$\\
 SRT-P-01 & {\sc ottb} $kT=200$~keV & $[ -16, +100]$ &  1  & $2.3$  & $2.8$\\         
 SRT-P-01 & {\sc ottb} $kT=200$~keV & $[ -16, +100]$ & 16  & $4.1$  & $2.5$\\
 SRT-P-01 & {\sc ottb} $kT=200$~keV & $[ -16, +100]$ & 128 & $9.9$  & $3.0$\\
 SRT-P-02 & {\sc pl} $\Gamma=2$ & $[ -100, +100]$ &  1  & $2.0$  & $3.0$\\         
 SRT-P-02 & {\sc pl} $\Gamma=2$ & $[ -100, +100]$ & 16  & $3.5$  & $3.6$\\
 SRT-P-02 & {\sc pl} $\Gamma=2$ & $[ -100, +100]$ & 128 & $7.8$  & $2.9$\\
 SRT-P-02 & {\sc ottb} $kT=50$~keV & $[ -100, +100]$ &  1  & $2.8$  & $3.0$\\         
 SRT-P-02 & {\sc ottb} $kT=50$~keV & $[ -100, +100]$ & 16  & $5.0$  & $3.6$\\
 SRT-P-02 & {\sc ottb} $kT=50$~keV & $[ -100, +100]$ & 128 & $11$  & $2.9$\\
 SRT-P-02 & {\sc ottb} $kT=200$~keV & $[ -100, +100]$ &  1  & $2.9$  & $3.0$\\ 
 SRT-P-02 & {\sc ottb} $kT=200$~keV & $[ -100, +100]$ & 16  & $5.2$  & $3.6$\\
 SRT-P-02 & {\sc ottb} $kT=200$~keV & $[ -100, +100]$ & 128 & $12$  & $2.9$\\
 SRT-P-07 & {\sc pl} $\Gamma=2$ & $[ -36, +100]$ &  1  & $1.6$  & $3.2$\\         
 SRT-P-07 & {\sc pl} $\Gamma=2$ & $[ -36, +100]$ & 16  & $2.7$  & $3.3$\\
 SRT-P-07 & {\sc pl} $\Gamma=2$ & $[ -36, +100]$ & 128 & $5.5$  & $3.0$\\
 SRT-P-07 & {\sc ottb} $kT=50$~keV & $[ -36, +100]$ &  1  & $2.2$  & $3.2$\\         
 SRT-P-07 & {\sc ottb} $kT=50$~keV & $[ -36, +100]$ & 16  & $3.9$  & $3.3$\\
 SRT-P-07 & {\sc ottb} $kT=50$~keV & $[ -36, +100]$ & 128 & $7.8$  & $3.0$\\
 SRT-P-07 & {\sc ottb} $kT=200$~keV & $[ -36, +100]$ &  1  & $2.3$  & $3.2$\\ 
 SRT-P-07 & {\sc ottb} $kT=200$~keV & $[ -36, +100]$ & 16  & $4.1$  & $3.3$\\
 SRT-P-07 & {\sc ottb} $kT=200$~keV & $[ -36, +100]$ & 128 & $8.2$  & $3.0$\\
 SRT-P-08 & {\sc pl} $\Gamma=2$ & $[ -51, +100]$ &  1  & $1.6$  & $3.1$\\         
 SRT-P-08 & {\sc pl} $\Gamma=2$ & $[ -51, +100]$ & 16  & $2.7$  & $3.2$\\
 SRT-P-08 & {\sc pl} $\Gamma=2$ & $[ -51, +100]$ & 128 & $5.5$  & $2.9$\\
 SRT-P-08 & {\sc ottb} $kT=50$~keV & $[ -51, +100]$ &  1  & $2.2$  & $3.1$\\         
 SRT-P-08 & {\sc ottb} $kT=50$~keV & $[ -51, +100]$ & 16  & $3.9$  & $3.2$\\
 SRT-P-08 & {\sc ottb} $kT=50$~keV & $[ -51, +100]$ & 128 & $7.8$  & $2.9$\\
 SRT-P-08 & {\sc ottb} $kT=200$~keV & $[ -51, +100]$ &  1  & $2.3$  & $3.1$\\ 
 SRT-P-08 & {\sc ottb} $kT=200$~keV & $[ -51, +100]$ & 16  & $4.1$  & $3.2$\\
 SRT-P-08 & {\sc ottb} $kT=200$~keV & $[ -51, +100]$ & 128 & $8.2$  & $2.9$\\ 
 SRT-P-09 & {\sc pl} $\Gamma=2$ & $[ -100, +100]$ &  1  & $1.6$  & $2.5$\\         
 SRT-P-09 & {\sc pl} $\Gamma=2$ & $[ -100, +100]$ & 16  & $3.1$  & $3.0$\\
 SRT-P-09 & {\sc pl} $\Gamma=2$ & $[ -100, +100]$ & 128 & $7.0$  & $2.9$\\
 SRT-P-09 & {\sc ottb} $kT=50$~keV & $[ -100, +100]$ &  1  & $2.2$  & $2.5$\\         
 SRT-P-09 & {\sc ottb} $kT=50$~keV & $[ -100, +100]$ & 16  & $4.5$  & $3.0$\\
 SRT-P-09 & {\sc ottb} $kT=50$~keV & $[ -100, +100]$ & 128 & $10$  & $2.9$\\
 SRT-P-09 & {\sc ottb} $kT=200$~keV & $[ -100, +100]$ &  1  & $2.3$  & $2.5$\\ 
 SRT-P-09 & {\sc ottb} $kT=200$~keV & $[ -100, +100]$ & 16  & $4.6$  & $3.0$\\
 SRT-P-09 & {\sc ottb} $kT=200$~keV & $[ -100, +100]$ & 128 & $11$  & $2.9$\\
 uGMRT-01 & {\sc pl} $\Gamma=2$ & $[ -100, +100]$ &  1  & $1.6 (3.6)$\tablefootmark{c}  & $2.7$\\
 uGMRT-01 & {\sc pl} $\Gamma=2$ & $[ -100, +100]$ & 16  & $3.1 (8.7)$\tablefootmark{c}  & $3.4$\\
 uGMRT-01 & {\sc pl} $\Gamma=2$ & $[ -100, +100]$ & 128 & $6.6 (30)$\tablefootmark{c}  & $3.3$\\
 uGMRT-01 & {\sc ottb} $kT=50$~keV & $[ -100, +100]$ &  1  & $2.2 (6.8)$\tablefootmark{c}  & $2.7$\\
 uGMRT-01 & {\sc ottb} $kT=50$~keV & $[ -100, +100]$ & 16  & $4.5 (16)$\tablefootmark{c}  & $3.4$\\
 uGMRT-01 & {\sc ottb} $kT=50$~keV & $[ -100, +100]$ & 128 & $9.5 (56)$\tablefootmark{c}  & $3.3$\\
 uGMRT-01 & {\sc ottb} $kT=200$~keV & $[ -100, +100]$ &  1  & $2.3 (12)$\tablefootmark{c}  & $2.7$\\ 
 uGMRT-01 & {\sc ottb} $kT=200$~keV & $[ -100, +100]$ & 16  & $4.6 (29)$\tablefootmark{c}  & $3.4$\\
 uGMRT-01 & {\sc ottb} $kT=200$~keV & $[ -100, +100]$ & 128 & $11 (97)$\tablefootmark{c}  & $3.3$ \\ 
 uGMRT-02 & {\sc pl} $\Gamma=2$ & $[ -100, +27.7]$ &  1  & $2.4$\tablefootmark{d}  & $2.6$\\
 uGMRT-02 & {\sc pl} $\Gamma=2$ & $[ -100, +27.7]$ & 16  & $4.7$\tablefootmark{d}  & $3.1$\\
 uGMRT-02 & {\sc pl} $\Gamma=2$ & $[ -100, +27.7]$ & 128 & $11.$\tablefootmark{d}  & $2.9$\\
 uGMRT-02 & {\sc ottb} $kT=50$~keV & $[ -100, +27.7]$ &  1  & $2.6$\tablefootmark{d}  & $2.6$\\
 uGMRT-02 & {\sc ottb} $kT=50$~keV & $[ -100, +27.7]$ & 16  & $5.1$\tablefootmark{d}  & $3.1$\\
 uGMRT-02 & {\sc ottb} $kT=50$~keV & $[ -100, +27.7]$ & 128 & $12.$\tablefootmark{d}  & $2.9$\\
 uGMRT-02 & {\sc ottb} $kT=200$~keV & $[ -100, +27.7]$ &  1  & $2.6$\tablefootmark{d}  & $2.6$\\ 
 uGMRT-02 & {\sc ottb} $kT=200$~keV & $[ -100, +27.7]$ & 16  & $5.3$\tablefootmark{d}  & $3.1$\\
 uGMRT-02 & {\sc ottb} $kT=200$~keV & $[ -100, +27.7]$ & 128 & $12$\tablefootmark{d}  & $2.9$ 
\end{longtable}
\tablefoot{
\tablefoottext{a}{{\sc pl}: power-law with photon index $\Gamma=2$; {\sc ottb}: optically thin thermal bremsstrahlung.}
\tablefoottext{b}{TDB since FRB time calculated at infinite frequency.}
\tablefoottext{c}{Values among parentheses refer to the 1--100~keV passband.}
\tablefoottext{d}{Values refer to the 10--30~keV passband.}
}
}

\end{appendix}

\end{document}